\documentclass[english,aps,reprint,superscriptaddress,longbibliography,floatfix]{revtex4-2}

\usepackage[T1]{fontenc}
\usepackage[latin9]{inputenc}
\usepackage{amssymb}
\usepackage{graphicx}
\usepackage{subscript}
\usepackage{amsmath}
\usepackage{booktabs}
\usepackage{graphicx}

\makeatletter
\usepackage{pslatex}

\usepackage{hyperref}% add hypertext capabilities
\usepackage{xcolor}
\hypersetup{
    colorlinks,
    linkcolor={blue!70!black},
    citecolor={blue!70!black},
    urlcolor={blue!70!black}
}

\usepackage{babel}

\makeatother

\begin{document}

\title{Enhanced sensitivity to ultralight bosonic dark matter in the spectra of the linear radical SrOH}

\author{Ivan Kozyryev}

\email{ikozyryev@gmail.com}

%\altaffiliation[Present address: ]{Department of Physics, Columbia University, New York, NY 10027}

\affiliation{Harvard-MIT Center for Ultracold Atoms, Cambridge, MA 02138}

\affiliation{Department of Physics, Harvard University, Cambridge, MA 02138}

\author{Zack Lasner}

\email{zlasner@g.harvard.edu }

\affiliation{Harvard-MIT Center for Ultracold Atoms, Cambridge, MA 02138}

\affiliation{Department of Physics, Harvard University, Cambridge, MA 02138}
%\affiliation{Department of Physics, Yale University, New Haven, CT 06511}

\author{John M. Doyle}

\affiliation{Harvard-MIT Center for Ultracold Atoms, Cambridge, MA 02138}

\affiliation{Department of Physics, Harvard University, Cambridge, MA 02138}

\date{\today}
\begin{abstract}
Coupling between Standard Model particles and theoretically well-motivated ultralight dark matter (UDM) candidates can lead to time variation of fundamental constants, including the proton-to-electron mass ratio $\mu\equiv m_{p}/m_{e}\approx1836$. The presence of nearly-degenerate vibrational energy levels of different character in polyatomic molecules can result in significantly enhanced relative energy shifts in molecular spectra originating from $\partial_{t}\mu$, relaxing experimental complexity required for high-sensitivity measurements. We analyze the amplification of UDM effects in the spectrum of laser-cooled strontium monohydroxide (SrOH). SrOH was the first polyatomic molecule to be directly laser cooled to sub-millikelvin temperatures [Kozyryev \emph{et al.}, Phys. Rev. Lett. \textbf{118}, 173201 (2017)], opening the possibility of long experimental coherence times and providing a promising platform for suppressing systematic errors. Because of the high enhancement factors ($\left|Q_{\mu}\right|\approx10^{3}$), measurements of the $\tilde{X}\left(200\right)\leftrightarrow\tilde{X}\left(03^{1}0\right)$ rovibrational transitions of SrOH in the microwave regime can result in $\sim10^{-17}$ fractional uncertainty in $\delta\mu/\mu$ with one day of integration, leading to significantly improved constraints for UDM coupling constants. We also detail how the use of more complex MOR-type radicals with additional vibrational modes arising from larger ligands R could lead to even greater enhancement factors, while still being susceptible to direct laser cooling. 
\end{abstract}
\maketitle
\section{Introduction}

The quantum mechanical nature of dark matter remains a mystery despite significant experimental efforts \cite{mitsou2015overview,baudis2012direct,CDMS2018results,PandaXII2016dark,LUX2017results}. Stringent limits placed recently on the promising class of dark matter candidates, Weakly Interacting Massive Particles \cite{baudis2014wimp,CDMS2018results}, as well as the absence of signatures for supersymmetric partners at the Large Hadron Collider \cite{CMS_dark_matter_2018,CMS_new_physics,mitsou2015overview} and electron electric dipole moment (EDM) experiments \cite{cairncross2017precision,ACME2014}, have motivated a new generation of searches for other theoretically motivated dark matter candidates \cite{graham2015cosmological,stadnik2017new,van2015search,irastorza2018new,berlin2018thermal,budker2017CASPEr,d2014multiverse}.
Bosonic ultralight dark matter (UDM) particles, like axions, axion-like particles (ALPs), dilatons, moduli, and relaxions \cite{graham2015cosmological}, can form coherently oscillating classical fields $\phi\left(\mathbf{r},\,t\right)=\phi_{0}\cos\left(\omega_{\phi}t-\mathbf{k_{\phi}\cdot}\mathbf{r}\right)$
with the oscillation frequency set by the mass of the dark matter particle $\omega_{\phi}\simeq m_{\phi}$ \cite{arvanitaki2015searching,stadnik2015can,guth2015dark}. Coupling between UDM fields and ordinary matter can lead to variation in fundamental constants $X=\alpha$ (fine-structure constant) and $\mu$ (proton-to-electron mass ratio) as \cite{roberts2018precision,stadnik2015can}
\begin{equation}
\frac{\delta X\left(t\right)}{X}=\Gamma_{X}\phi^{n}\left(\mathbf{r},t\right),
\end{equation}
where the coupling strength is $\Gamma_{X}$ and $n=1(2)$ for linear(quadratic) coupling.

Transitions between different quantum levels with energy separation $\triangle E=\hbar\omega$ in atoms and molecules are dependent on the dimensionless constants $\delta\omega=f\left(\delta\alpha,\delta\mu\right)$ with \cite{kozlov2013microwave,jansen2014perspective} 
\begin{equation}
\frac{\delta\omega}{\omega}=Q_{\alpha}\frac{\delta\alpha}{\alpha}+Q_{\mu}\frac{\delta\mu}{\mu}.
\end{equation}
Sensitive probes of $\alpha$ variation due to UDM-induced effects have recently been explored with the use of ultraprecise atomic clocks \cite{arvanitaki2015searching,dzuba2018additional,safronova2018two}, reaching $\partial_{t}\alpha/\alpha\sim10^{-17}$/yr sensitivity \cite{godun2014frequency,ytterbium2014improved,rosenband2008frequency}. Additionally, specific atomic transitions with enhanced sensitivities $\left|Q_{\alpha}\right|\gg1$ have allowed measurements on a Dy beam to be competitive with atomic clock limits \cite{van2015search,leefer2013new}. Exploring both $\Gamma_{\mu}$ and $\Gamma_{\alpha}$ is important as these effects probe different underlying physical phenomena \cite{arvanitaki2015searching}. While the use of atomic clocks for probing dark-matter-induced oscillating, drifting and transient-in-time fundamental constants has been considered in depth \cite{roberts2017search,roberts2018precision,wcislo2016searching}, laser-cooled molecules have additional degrees of freedom that could enable further breakthroughs in this area.

In molecular spectra, the energy scales for electronic, vibrational and rotational transitions typically relate as $1:\mu^{-1/2}:\mu^{-1}$ \cite{demille2015diatomic}. Molecular transitions provide a system to study $\Gamma_{\mu}$ couplings without any contributions from $\Gamma_{\alpha}$ because vibrational transitions in molecules have $Q_{\mu}=-\frac{1}{2}$ and $Q_{\alpha}\approx0$ \cite{jansen2014perspective}. Thus, isolating effects from $\mu$ variation in a model-independent manner becomes possible \cite{shelkovnikov2008stability}. Moreover, certain beyond the Standard Model theories predict larger $\mu$ variation $\delta\mu/\mu=R\delta\alpha/\alpha$ with $R\approx40$ \cite{jansen2014perspective,flambaum2004limits,calmet2002cosmological}, further motivating precision experiments in molecular spectroscopy. Molecular ions can also be used for such experiments and recent theoretical proposals consider using diatomic hetero- and homo-nuclear molecular ions to search for $\mu$ variation \cite{hanneke2016high,kokish2018prospects}. In this paper, we propose to use laser-cooled samples of the neutral polyatomic radical SrOH that can be trapped at high densities and low temperatures, allowing for large scalability and enhanced sensitivity to UDM-induced $\mu$ variation.

\section{Enhanced sensitivity to dark matter with near-degenerate states}\label{sec:enhancement-factors}

\begin{figure}
\begin{centering}
\includegraphics[width=6cm]{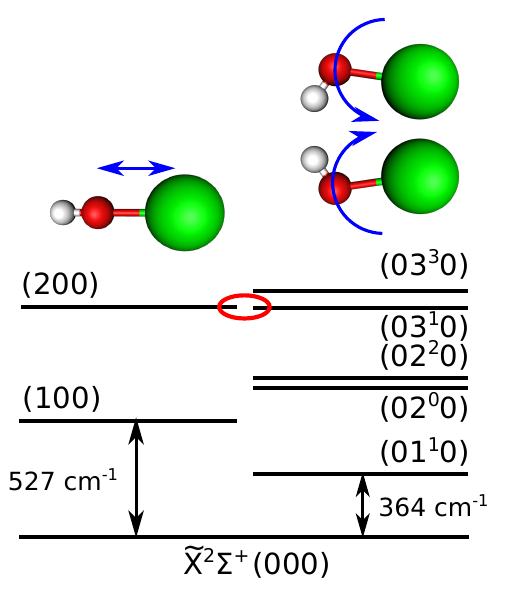} 
\par\end{centering}
\caption{\label{fig:Vibrational-levels-of-SrOH}Vibrational levels of SrOH in the electronic ground state. The dominant nature of the vibrational motion is indicated schematically at the top. The states are labeled using $\left(v_{1}v_{2}^{l}v_{3}\right)$ notation for the number of vibrational quanta in the Sr-O stretching ($v_{1}$), Sr-O-H bending ($v_{2}$) and O-H stretching ($v_{3}$) vibrational motion. The number of units of vibrational angular momentum present in the doubly-degenerate bending of linear triatomic molecules is denoted with the superscript $l$. Nearly-degenerate excited stretching $\left(200\right)$ and bending $\left(03^{1}0\right)$ vibrations are indicated with a red oval. }
\end{figure}

As previously pointed out \cite{demille2008enhanced,flambaum2007enhanced,zelevinsky2008precision,kozlov2013linear,kozlov2013microwave}, rovibrational spectroscopy of diatomic and polyatomic molecules may provide significant enhancements in relative sensitivity to the variation of $\mu$ with $\left|Q_{\mu}\right|\gg1$. An extensive list of enhancement factors calculated for diatomic and polyatomic molecules to $\mu$ variation can be found in Refs. \cite{jansen2014perspective,kozlov2013microwave}, with large enhancements of $Q_{\mu}\sim300$ and $Q_{\mu}\sim700$ estimated for CH\textsubscript{3}OH and $l$-C\textsubscript{3}H, respectively. Other polyatomic molecules found in space like methanol \cite{jansen2011methanol}, acetone \cite{ilyushin2014sensitivity} and ammonia \cite{owens2016enhanced} have been analyzed as well, leading to a stringent limit of $\delta\mu/\mu\lesssim10^{-7}$ from the observations of astronomical methanol \cite{bagdonaite2012stringent,jansen2014perspective}. While astrophysical observations place stringent time-variation limits with $\partial_{t}\mu/\mu\sim10^{-17}$/yr bounds due to large look-back times ($\triangle t\sim7\,{\rm Gyr}$) \cite{jansen2014perspective}, they have limited sensitivity to UDM-induced coherent oscillations since a linear drift, $\partial_{t}\mu=\delta\mu/\triangle t$, must be assumed. 

Here we analyze the enhancement factors for one of the simplest possible polyatomic molecules, the linear triatomic XYZ-type radical SrOH, and discover that enhancement factors of $Q_{\mu}\approx10-10^{3}$ can be reached by probing rovibrational transitions of the $\tilde{X}\left(200\right)\leftrightarrow\tilde{X}\left(03^{1}0\right)$ excitation spectrum in the $\omega\approx2\pi\times1-30$ GHz transition frequency band. SrOH was the first polyatomic molecule to be directly laser cooled \cite{kozyryev2017sisyphus}, and thus provides the additional significant advantages of low translational and internal temperatures, long experimental coherence times, and optical internal state preparation and efficient readout. Furthermore, the simple vibrational structure of SrOH strongly limits the possibility of internal vibrational redistribution (IVR) or nonradiative transitions \cite{Demtroeder2005}, enabling highly sensitive laboratory measurements of both $\delta\mu/\mu$ and $\partial_{t}\mu/\mu$ in the frequency band of theoretical interest for promising UDM models. Figure \ref{fig:Vibrational-levels-of-SrOH} shows the relevant vibrational energy levels of SrOH in the ground electronic state $\tilde{X}$.

To see how a large sensitivity to $\mu$ variation arises in the rovibrational spectrum of SrOH, we begin by following previous treatments in Refs. \cite{kozlov2013microwave,jansen2014perspective}. Consider two different energy levels $E_{g}$ and $E_{e}$ within the same electronic state with $E_{g}<E_{e}$. The energy difference is (assuming $\hbar=1$) 
\begin{equation}
\omega=E_{e}-E_{g}\label{eq:omega_def}
\end{equation}
with the change arising from $\mu$ variation given as 
\begin{equation}
\delta\omega=\frac{\partial E_{e}}{\partial\mu}\delta\mu-\frac{\partial E_{g}}{\partial\mu}\delta\mu.
\end{equation}
Therefore, the fractional change in the level separation is 
\begin{equation}
\frac{\delta\omega}{\omega}=\frac{1}{E_{e}-E_{g}}\left(\frac{\partial E_{e}}{\partial\mu}-\frac{\partial E_{g}}{\partial\mu}\right)\delta\mu.
\end{equation}
Equivalently, the relationship between the fractional changes in $\omega$ and $\mu$ are related to each other as 
\begin{equation}
\frac{\delta\omega}{\omega}=Q_{\mu}\frac{\delta\mu}{\mu}
\end{equation}
with the proportionality constant $Q_{\mu}$ also known as the dimensionless enhancement factor defined as 
\begin{equation}
Q_{\mu}\equiv\frac{\mu}{E_{e}-E_{g}}\left(\frac{\partial E_{e}}{\partial\mu}-\frac{\partial E_{g}}{\partial\mu}\right)
\end{equation}
or as more common in the literature 
\begin{equation}
Q_{\mu}\equiv\frac{1}{\omega}\left(\frac{\partial E_{e}}{\partial\left(\ln\mu\right)}-\frac{\partial E_{g}}{\partial\left(\ln\mu\right)}\right).\label{eq:big_Q}
\end{equation}
The absolute dependence of each energy level is calculated as 
\begin{equation}
q_{g,e}\equiv\frac{\partial E_{g,e}}{\partial\left(\ln\mu\right)}
\end{equation}
and has units of energy (usually cm\textsuperscript{-1}). From Eq. \ref{eq:big_Q} one can observe that a large enhancement factor $Q_{\mu}$ will arise when two levels being probed are closely spaced (i.e. $\omega\approx0$) and have different dependence on $\mu$ (i.e. $q_{g}\neq q_{e}$).

Generically, the interplay between harmonic and anharmonic contributions (discussed in detail for SrOH below) to the difference in sensitivity coefficients, $\triangle q$, can lead to enhancement factors $Q_{\mu}$ significantly larger than unity. In order to demonstrate the role of both harmonic and anharmonic terms, we consider two vibrational levels separated by $\triangle E=\triangle E_{{\rm harm}}+\triangle E_{{\rm anharm}}$. Using the dependence of vibrational constants on the proton-to-electron mass ratio (see App.~\ref{sec:anharmonic-vibrations}) we calculate $\triangle q=-\frac{1}{2}\triangle E_{{\rm harm}}-\triangle E_{{\rm anharm}}$ for the $\mu$ sensitivity difference. Therefore, the absolute enhancement factor $Q_{\mu}=\triangle q/\triangle E$ becomes
\begin{equation}\label{eq:Qmu-anharm-harm}
Q_{\mu}=-\frac{1}{2}\left(1+\frac{\Delta E_{{\rm anharm}}}{\triangle E_{{\rm harm}}+\triangle E_{{\rm anharm}}}\right).
\end{equation}
For illustration, we consider three limiting cases, depending on the relative contributions of $\triangle E_{{\rm harm}}$ and $\triangle E_{{\rm anharm}}$:
\begin{align}
\triangle E_{{\rm anharm}}=0:\,Q_{\mu}\rightarrow-\frac{1}{2}\nonumber \\
\triangle E_{{\rm harm}}=0:\,Q_{\mu}\rightarrow-1\\
\triangle E_{{\rm harm}}\sim-\triangle E_{{\rm anharm}}:\,Q_{\mu}\rightarrow\pm\infty.\nonumber 
\end{align}
Therefore, a large enhancement factor is expected for a transition with anharmonic contributions comparable in magnitude to the harmonic oscillator energy difference and opposite in sign. Inclusion of the small rotational energy difference $\triangle E_{{\rm rot}}$ in a given rovibrational transition leaves Eq.~\ref{eq:Qmu-anharm-harm} unchanged up to the substitution $\triangle E_{{\rm anharm}} \rightarrow \triangle E_{{\rm anharm}} + \triangle E_{{\rm rot}}$.

% For the $\left(200\right)\leftrightarrow\left(03^{1}0\right)$ vibronic transition in SrOH, $\triangle E_{{\rm harm}}=2\omega_{1}-3\omega_{2}\approx-44.951$ cm\textsuperscript{-1} and $\triangle E_{{\rm anharm}}=6x_{11}-15x_{22}-g_{22}\approx44.99$ cm\textsuperscript{-1}, leading to $\triangle E_{{\rm harm}}/\triangle E_{{\rm anharm}}\approx-0.999122$ and $Q_{\mu}\approx-570$.

We now consider the harmonic and anharmonic contributions to the energy of rovibrational states in SrOH. As shown in App.~\ref{sec:molecular-constants}, for a linear triatomic molecule like SrOH, the positions of the vibrational energy levels $\left(v_{1}v_{2}^{l}v_{3}\right)$ referenced relative to the lowest level $\left(000\right)$ in a given electronic state are described as \cite{herzberg1966polyatomics} 
\begin{equation}
E_{v_{1}v_{2}v_{3}-000}=\sum_{i=1}^{3}\left(\omega_{i}v_{i}+x_{ii}v_{i}^{2}+x_{ii}v_{i}d_{i}\right)+g_{22}l_{2}^{2}
\end{equation}
with $d_{i}=1$ for the stretching modes with frequencies $\omega_{1}$ (Sr$\leftrightarrow$O) and $\omega_{3}$ (O$\leftrightarrow$H), and $d_{i}=2$ for the doubly-degenerate bending mode with frequency $\omega_{2}$. The anharmonic contributions to the molecular potentials have been included leading to additional $x_{ii}$ and $g_{ii}$ terms in the expansion. The expressions for the two closely-lying vibrational levels of SrOH shown in Fig. \ref{fig:Vibrational-levels-of-SrOH} are given as $E_{200-000}=2\omega_{1}+6x_{11}$ and $E_{03^{1}0-000}=3\omega_{2}+15x_{22}+g_{22}$. With the estimated molecular constants for SrOH based on experimental measurements \cite{presunka1995laser}, we determine the energy separation between the two states $\triangle E_{200-03^{1}0}$ to be 
\begin{equation}
2\omega_{1}+6x_{11}-3\omega_{2}-15x_{22}-g_{22}=0.0395\,{\rm cm^{-1}},
\end{equation}
which corresponds to about 1.2 GHz. As discussed above, because the harmonic and anharmonic contributions to $\Delta E_{200-03^{1}0}=\Delta E_{{\rm harm}}+\Delta E_{{\rm anharm}}$ depend differently on $\mu$, the transition frequency displays a strong sensitivity to $\mu$ that is not suppressed even in the limit of degeneracy. In this regime, extremely small absolute energy shifts, $\delta\Delta E_{200-03^{1}0}<10\,\mu{\rm Hz}$, can be experimentally resolved, providing a sensitive probe of $\delta\mu\propto\delta\Delta E_{200-03^{1}0}$.

While the dominant energy scale arises from vibration, the smaller contribution to $Q_\mu$ from rotational motion becomes important when the vibrational energies between two states are nearly degenerate. For the ground electronic state $\tilde{X}$ of SrOH, the valence electron is effectively localized on the Sr atom and the unpaired electron spin is not strongly bound to the internuclear molecular symmetry axis $z$ \cite{brazier1985laser}. Therefore, rotational levels in both $\left(200\right)$ and $\left(03^{1}0\right)$ vibrational states can be analyzed in terms of Hund's coupling case (b) quantum numbers \cite{herzberg1966polyatomics} as $F_{\left[v\right]}\left(N\right)=B_{\left[v\right]}N\left(N+1\right)$, where $N$ is the quantum number of the total angular momentum apart from spin and $B_{\left[v\right]}$ is a rotational constant for a specific vibrational level $\left[v\right]$.

Using the dependence of the harmonic ($\omega_{i}$), anharmonic ($x_{ii}$, $g_{ii}$) and rotational ($B_{i}$) coefficients on the proton-to-electron mass ratio $\mu$ \cite{jansen2014perspective}, we calculate the absolute sensitivity of each rovibrational level $\left[N,v\right]$ to be
\begin{equation}
q_{[N,v]}\equiv\frac{\partial E_{[N,v]}}{\partial\left(\ln\mu\right)}=-\frac{1}{2}\omega_{i}v-x_{ii}\left(v^{2}+vd_{i}\right)
\end{equation}
\[
-g_{ii}l_{i}^{2}-B_{\left[v\right]}N\left(N+1\right),
\]
where $l_{1}(l_{2})=0(1)$, $d_{1}(d_{2})=1(2)$ and the sensitivity $q_{[0,000]}$ of the ground vibrational level has been subtracted.
% Using spectroscopic data for SrOH (see App.~\ref{sec:molecular-constants}) we obtain the absolute sensitivity coefficients $q_{200}\left(N=1\right)\approx-518\,{\rm cm^{-1}}$ and $q_{03^{1}0}\left(N=1\right)\approx-495\,{\rm cm^{-1}}$, which can be used to estimate the dimensionless enhancement factor as $Q_{\mu}=-617$ with transition frequency $\omega=2\pi\times 1.1$ GHz.
Each of the rotational levels in the $\left(03^{1}0\right)$ vibrational state consists of $\ell$-type parity doublets separated by $\Delta E_{\pm l}\sim\mathcal{O}\left(B_{\left[v\right]}^{2}/\omega_{2}\right)$ which has been measured for SrOH in this specific vibrational level to be $\Delta E_{\pm l}\approx12$ MHz \cite{fletcher1995millimeter}. Driving the perpendicular vibronic transition $\Sigma-\Pi$ with $\triangle l=\pm1$ leads to $P$ and $R$ branches with $\triangle N=\pm1$, as well as a strong $Q$ branch with $\triangle N=0$ \cite{Bernath2005}. The relative sensitivity coefficient of the rovibrational $N''=1\rightarrow N'=1$ transition for $\tilde{X}\left(200\right)\rightarrow\tilde{X}\left(03^{1}0\right)$ is estimated to be $Q_{\mu}=-617$ with transition frequency $\omega=2\pi\times1.1$ GHz. By choosing the $N''=2\rightarrow N'=1$ rotational branch instead, we obtain $Q_{\mu}=-23$ with $\omega=2\pi\times31$ GHz. The sign of the shift can be reversed by using the other transition branch $N''=1\rightarrow N'=2$ with $Q_{\mu}=23$ and $\omega=2\pi\times29$ GHz. Thus, by measuring different rotational branches of the same vibrational transition the sign and magnitude of the sensitivity enhancement factor $Q_{\mu}$ can be controlled. Vibrational dependence of the rotational constant $B_{\left[v\right]}$ can be used to achieve even larger $Q_{\mu}$ since $\triangle B_{200-03^{1}0}\approx-45$ MHz \cite{fletcher1995millimeter,presunka1995laser}. For the $N''=5\rightarrow N'=5$ rotational branch, the separation between the levels is estimated to decrease to $\triangle E_{200-03^{1}0}<200$ MHz, resulting in $Q_{\mu}>10^{3}$ enhancement (see Sec.~\ref{sec:enhancement-uncertainty} for discussion of the uncertainty in these estimates). As a stability reference, one could use purely rotational transitions within the $\left(200\right)$ vibrational manifold with $Q_{\mu}=-1$. It is important to note that our spectroscopic constants derived from previous experimental measurements \cite{presunka1995laser} reproduce positions of $E_{100}$, $E_{200}$, $E_{01^{1}0}$, $E_{02^{0}0}$ and $E_{02^{2}0}$ to within 0.002 cm\textsuperscript{-1} (see App.~\ref{sec:molecular-constants}). Furthermore, the absolute magnitude of the calculated enhancement factors $Q_{\mu}$ is comparable to the largest values found in the literature for much more complex polyatomic molecules like methanol \cite{levshakov2011methanol} and ammonia \cite{owens2016enhanced}.

\subsection{Enhancement factor uncertainty}\label{sec:enhancement-uncertainty}

In our analysis of the anharmonic contributions to the vibrational potential of SrOH, we have ignored the terms arising from coupling between different vibrational modes (i.e. $x_{ij}$ with $i\neq j$) in Eq. \ref{eq:Vibrational_term_value}. While the vibrational potential for SrOH is mostly harmonic with $\omega_{i}\gg x_{ii},\,x_{ij}$, contributions from the $x_{ij}$ terms could lead to shifts on the order of a few cm\textsuperscript{-1}. Previous experimental bounds on the location of the $\left(03^{3}0\right)$ vibrational level along with the estimate of the $g_{22}$ coefficient further confirm that $\triangle E_{03^{1}0-200}\lesssim2$ cm\textsuperscript{-1} \cite{presunka1995laser}. While exact spectroscopy of the $\left(03^{1}0\right)$ level in reference to a known vibronic level is necessary to determine the separation between $\left(200\right)$ and $\left(03^{1}0\right)$, we estimate an absolute worst-case value of $|Q_{\mu}|\approx100$. For a generic value of $\triangle E_{03^{1}0-200}<2$ cm\textsuperscript{-1}, we can identify a new optimal pair of rotational levels to use in a $P$ or $R$ branch as
\begin{equation}
\triangle E_{03^{1}0-200}\approx-\Delta E_{{\rm rot}}=B\left(N\left(N+1\right)-\left(N-1\right)N\right)
\end{equation}
\begin{equation}
\Rightarrow N\approx\frac{\triangle E_{03^{1}0-200}}{2B}\leq\frac{2\,{\rm cm^{-1}}}{2\times0.25\:{\rm cm^{-1}}}=4
\end{equation}
where $\Delta E_{\rm rot}$ is the difference in rotational energies and we used for the rotational constants $B\equiv B_{200}\approx B_{03^{1}0}\approx0.25\,{\rm cm^{-1}}$ \cite{presunka1995laser,fletcher1995excited}. In the worst case, the total angular transition frequency $\omega=|\Delta E_{03^{1}0-200}+\Delta E_{\rm rot}|$ cannot be made smaller than $B$. Therefore, $\omega<2\pi\times7.5$ GHz and $|Q_{\mu}|>100$. In a typical (rather than worst-case) scenario, enhancement factors significantly larger than this limit would be achieved. Thus, comparable sensitivity can be reached as estimated using the best-fit spectroscopic constants currently available.

\section{Sensitivity estimation}

In addition to the large relative enhancement factors to $\mu$ value variation, SrOH uniquely provides an intriguing experimental platform for achieving precise measurements of $\delta\mu/\mu$ using previously demonstrated atomic physics technologies. For atomic clock experiments, the statistical precision with which the transition frequency can be measured, with the frequency stability limited by quantum projection noise, is $\delta\omega\approx1/\sqrt{NT_{c}\tau}$ \cite{ludlow2015optical,hanneke2016high}, where $N$ is the number of independent molecules probed per run, $T_{c}$ is the experimental coherence time, and $\tau$ is the total measurement time. Vibrational motions of SrOH are quite harmonic for low quantum numbers and, therefore, radiative vibrational decays with $\triangle v\neq\pm1$ are suppressed. Thus, the coherence time in the experiment $T_{c}$ will be limited by the spontaneous vibrational decay from $\tilde{X}\left(200\right)$ to $\tilde{X}\left(100\right)$, which we estimate to be $\sim140$ ms (see Sec.~\ref{sec:lifetime-estimate}). Black-body stimulated lifetime at room temperature is estimated to be $T_{{\rm BBR}}>1.5$ s, consistent with previous theoretical estimates \cite{vanhaecke2007precision}.

Exploiting the full coherence time of the $\tilde{X}(200)-\tilde{X}(03^10)$ system requires laser cooling and trapping SrOH molecules. Direct laser cooling of $10^6$ SrOH molecules to millikelvin temperatures has already been demonstrated \cite{kozyryev2017sisyphus}. With the Doppler cooling technique, which relies on the spontaneous radiation pressure force, the transverse temperature of a cryogenic SrOH beam was reduced to 30 mK \cite{KozyryevThesis}. Additionally, the use of the sub-Doppler cooling method known as magnetically-assisted Sisyphus laser cooling reduced the temperature to $\sim750\,{\rm \mu K}$ \cite{kozyryev2017sisyphus}. Detailed measurements of Franck-Condon factors (FCFs) and vibrational branching ratios (VBRs) for SrOH have been completed \cite{nguyen2018fluorescence}, confirming that direct laser slowing and magneto-optical trapping appears feasible predominantly with three repumping lasers to address losses to the $\left(100\right)$, $\left(200\right)$, and $\left(02^{0}0\right)$ states. Potentially, even fewer repumping lasers could be used employing slowing with coherent stimulated optical forces recently experimentally demonstrated for SrOH \cite{kozyryev2018coherent}. Sympathetic cooling of trapped SrOH to microkelvin temperatures with ultracold lithium also appears feasible based on rigorous quantum scattering calculations \cite{morita2017cold}. Direct magneto-optical trapping of $\sim10^{6}$ diatomic CaF molecules has already been demonstrated \cite{anderegg2017radio}.

With a combination of these demonstrated techniques, it is realistic to assume $N\approx10^{6}$ trapped SrOH molecules per experimental run. Long coherence times with laser-cooled SrOH molecules can be realized utilizing either an optical dipole trap or a molecular fountain \cite{anderegg2018laser,cheng2016molecular}. The required experimental coherence time $T_{c}$ is a factor of 5 shorter than the achieved lifetime of laser-cooled CaF in a red-detuned optical dipole trap \cite{anderegg2018laser}. Alternatively, a blue-detuned ``box'' trap \cite{chu1995long} would enable similarly long trap times. Precision spectroscopy of laser-cooled atomic radium has previously been performed in an optical dipole trap \cite{radium2015first}, demonstrating the feasibility of the optical approach.

With $10^6$ trapped molecules per experimental cycle, repeated every $T_{c}$, and one day of experimental integration, an absolute statistical uncertainty of $\delta\omega\approx10\,{\rm \mu Hz}$ can be achieved. Enhanced sensitivity coefficients $Q_{\mu}$ in SrOH spectra provide an opportunity to perform sensitive measurements with relaxed experimental precision, similar to gains in $\alpha$ variation sensitivity for Dy experiments \cite{leefer2013new}. The frequency of the rotational transitions addressed during the experiment on the $\tilde{X}\left(200\right)\leftrightarrow\tilde{X}\left(03^{1}0\right)$ vibrational band ranges between 1 and 30 GHz, and therefore expected relative measurement uncertainty is between $3\times10^{-12}/\sqrt{\tau\left({\rm seconds}\right)}$ and $1\times10^{-13}/\sqrt{\tau\left({\rm seconds}\right)}$. For comparison, microwave frequency synthesizers in the comparable frequency range $\omega\sim2\pi\times10\,{\rm GHz}$ for use in atomic clock experiments have microhertz resolution and noise levels at the $10^{-14}/\sqrt{\tau{\rm (seconds)}}$ level \cite{li2018low}.

Combining this expected frequency precision with the enhancement factors estimated in Sec.~\ref{sec:enhancement-factors}, we can achieve a fractional sensitivity $\delta\mu/\mu$ on the order of $\sim1\times10^{-17}$ for both $\omega\approx2\pi\times1$ and $2\pi\times30$ GHz transition frequencies (for a detailed discussion of the frequency-dependent sensitivity under different measurement scenarios, see Sec.~\ref{sec:cosmic-fields}). Thus, microwave spectroscopy of SrOH can provide $\delta\mu/\mu$ sensitivity at the level of the best previously proposed ultracold atom and trapped diatomic neutral \cite{flambaum2007enhanced,demille2008enhanced,zelevinsky2008precision} and ionic species \cite{hanneke2016high,kokish2018prospects}, but with potentially easier experimental preparation and spectroscopy schemes, as well as suppression of systematic errors as described in Sec.~\ref{sec:systematics}. Furthermore, the measurement with SrOH would lead to orders of magnitude improvement in the limit on $\mu$ variation in a model independent way compared to the previous experimental results with SF\textsubscript{6} beam spectroscopy \cite{shelkovnikov2008stability} or photoassociated ultracold KRb molecules where $\lesssim10^{-14}$/yr sensitivity was achieved \cite{kobayashi2019measurement}.

\section{Experimental details}

In this section, we show in greater detail the feasibility of transferring population to one of the nearly-degenerate vibrational states, driving the nominally forbidden microwave transition, and achieving long vibrational coherence times.

\subsection{State preparation}

\begin{figure}
\centering{}\includegraphics[width=8cm]{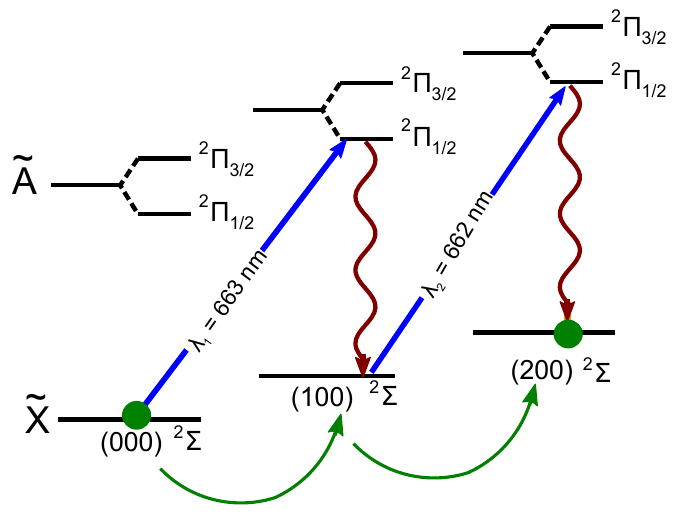}\caption{\label{fig:Internal-quantum-state-prep}Internal quantum state preparation
for SrOH via two-stage optical pumping (OP). Using two OP laser beams
($\lambda_{1}$ and $\lambda_{2}$), a trapped SrOH sample can be
prepared in a specific rotational quantum level of the excited Sr-O
stretching vibrational level (200). Spin-orbit splitting in the excited
electronic state is indicated with both $\Pi_{1/2}$ and $\Pi_{3/2}$
levels shown. }
\end{figure}

The efficient preparation of the necessary rovibrational quantum states can be achieved in two distinct ways. First, a two-stage optical pumping scheme from the ground vibrational level can populate $\left(200\right)$ via two stages of excitation to vibrationally excited levels of the $\tilde{A}^2\Pi_{1/2}$ electronic state (see Fig.~\ref{fig:Internal-quantum-state-prep}). In the first stage, molecules would be excited to $\tilde{A}^2\Pi_{1/2}(100)$, which efficiently decays to $\tilde{X}(100)$. That state, in turn, could be excited to $\tilde{A}(200)$, which would likewise preferentially decay to $\tilde{X}(200)$. Previous work on collisional quenching of the $\tilde{X}\left(100\right)$ state of SrOH at 2 K has already demonstrated high-efficiency optical pumping into the excited Sr-O stretching mode with a 660 nm external cavity diode laser \cite{kozyryev2015collisional}. Thus, efficient rotational state preparation in the $\left(200\right)$ state can be achieved with two optical pumping beams.

An alternative transfer scheme from the ground vibrational level to the excited $\tilde{X}\left(200\right)$ state is to turn off the $\tilde{X}\left(200\right)\rightarrow\tilde{A}^{2}\Pi_{1/2}\left(100\right)$ repumping laser during the laser cooling process, thus leading to the rapid accumulation of molecules in the $\tilde{X}\left(200\right)$ vibrational level. Each of the proposed methods appears highly feasible and the exact requirements of the future experiment will determine the preferred internal transfer scheme.

\subsection{Transition strength}

For linear molecules the intensity of rovibrational transitions within the same electronic state is estimated as $S_{J'J''}=\left|\mathbf{M}_{v'v''}\right|^{2}S_{J''}^{\triangle J}F\left(m\right)$, where $\mathbf{M}_{v'v''}$ represents a purely vibrational transition moment, $S_{J''}^{\triangle J}$ is the H\"onl-London factor and $F\left(m\right)$ is the Herman-Wallis term that compensates for errors in separation of vibration from rotation \cite{Bernath2005}. While for a purely harmonic oscillator only $\triangle v=\pm1$ transitions are allowed, inclusion of anharmonic terms in the molecular vibrational potential as well as high-order terms in the dipole moment function lead to overtones of reasonable intensity with $\triangle v=\pm2,\,\pm3,\ldots$ \cite{Bernath2005}. Additionally, for polyatomic molecules with nearby vibrational levels of different symmetry character (e.g. $\Sigma$ vs $\Pi$) like SrOH, Coriolis perturbations lead to Coriolis resonances and mixing between levels. The $\left(200\right)\sim\left(03^{1}0\right)$ Coriolis interaction for SrOH has been suggested previously \cite{presunka1995laser}. Combination transitions requiring changes in multiple $v$ quanta induced by the Coriolis interactions have previously been observed in other polyatomic molecules \cite{Field2000observation}.

% In the harmonic oscillator approximation to the vibrational motion, only $\Delta v=\pm1$ transitions are allowed. However, both anharmonic contributions to the vibrational potential and energy and Coriolis interactions that couple rotational and vibrational degrees of freedom can induce overtone and combination band transitions.

To quantitatively estimate the vibrational transition moment between (200) and ($03^10$), we consider here the interactions that induce a strong transition dipole moment between $(200)$ and $(03^10)$ in SrOH. By the symmetry of a linear molecule, anharmonic perturbations must be even in the bending normal coordinate $Q_2$ and therefore can't change $v_2$ by an odd number. Likewise, Coriolis interactions change $v_1+v_2$ by an even number at all orders of perturbation theory. Thus neither anharmonic, nor Coriolis, effects alone can induce a transition with $\triangle v_1 = 2$ and $\triangle v_2 = 3$. However, a combination of anharmonic and Coriolis interactions lead to a relatively strong transition between $(200)$ and $(03^10)$.

Matrix elements for the Coriolis interaction couple $v_1$ and $v_2$, and may be found in \cite{merer1971rotational}. Their strength is characterized by the Coriolis coefficient $\zeta_{21}$, which depends only on the atomic masses and geometry of a molecule \cite{meal1956vibration-rotation}. For SrOH, we find $\zeta_{21}=0.98$.

The vibrational potential energy for a linear polyatomic molecule expanded in terms of the dimensionless normal coordinates $q_{i}=Q_{i}\sqrt{2\pi c\omega_{i}/\hbar}$ is 
\begin{equation}%\label{eq:potential-energy}
V/hc=\frac{1}{2}\sum_{i}\omega_{i}q_{i}^{2}+\frac{1}{6}\sum_{ijk}\phi_{ijk}q_{i}q_{j}q_{k}+\frac{1}{24}\sum_{ijkl}\phi_{ijkl}q_{i}q_{j}q_{k}q_{l}+\ldots
\label{eq:Vanharmonic}\end{equation}
where $\text{\ensuremath{\phi}}_{ijk}$ and $\phi_{ijkl}$ are the cubic and quartic anharmonic force constants, respectively \cite{allen1990systematic}. We use force constants up to quartic order, computed from the potential energy surface (PES) calculation in \cite{li2019emulating}. As has been observed for CaOH \cite{koput2002ab}, the term $\frac{1}{6}\phi_{122}q_1q_2^2$ cannot be treated perturbatively due to vibrational near-resonances; we therefore directly diagonalize the Hamiltonian including the full vibrational energy with anharmonic terms $\phi_{ijk}$ and $\phi_{ijkl}$, as well as the Coriolis interaction. Our numerical results give $E_{100-000}=515$ cm$^{-1}$ and $E_{01^10-000}=333$ cm$^{-1}$, agreeing with experimental observations to better than 10\%. As expected, the $(200)$ and $(03^10)$ are found to be degenerate within the $10\%$ estimated uncertainty of the \emph{ab initio} energies.

Diagonalizing the Hamiltonian produces a set of vibrational eigenstates $|\psi_{v_1v_2^lv_3}\rangle$ expanded in terms of the harmonic oscillator basis, where the subscript labels the predominant basis component of the state. We then compute the transition dipole moment as $\sum_{v',v''}\langle\psi_{200}|\mathbf{M}_{v'v''}|\psi_{03^10}\rangle$, where here $\mathbf{M}_{v'v''}$ gives the characteristic transition strength between hypothetical pure harmonic oscillator states. Following the discussion in Sec.~\ref{sec:lifetime-estimate}, we estimate that $\mathbf{M}_{v'v''}=0.4$ D for stretching mode transitions in the harmonic oscillator basis, i.e. where $\Delta v_1=\pm1$ and $\Delta v_2=0$. Likewise, we estimate that $\mathbf{M}_{v'v''}=0.1$ D for bending mode transitions in the harmonic oscillator basis, with $\Delta v_2=\pm1$ and $\Delta v_1=0$ (see \cite{kozyryev2017precision}).

The resulting vibrational transition dipole moment for $(200)\leftrightarrow(03^10)$ is estimated to be in the range $0.02 - 0.04$ D, depending slightly on the specific rotational transition considered due to the $J$-dependence of the Coriolis interaction. This compares favorably with other proposed measurements, which typically rely on transition dipole moments of order $\leq 0.01$ D \cite{demille2008enhanced,flambaum2007enhanced,hanneke2016high,kobayashi2019measurement,kokish2018prospects}.

\subsection{Estimation of vibrational lifetime}\label{sec:lifetime-estimate}

The coherence time in the experiment will be limited by the spontaneous vibrational lifetime of the $\left(200\right)$ vibrational state. Specifically, the decay rate $\tilde{X}\left(200\right)\rightarrow\tilde{X}\left(100\right)$ can be estimated as $A_{200-100}=3.136\times10^{-7}\tilde{\omega}^{3}\mathbf{M}_{200-100}^{2}$ where $\tilde{\omega}=522$ is the energy splitting in cm\textsuperscript{-1} and the transition dipole moment $\mathbf{M}_{200-100}\approx0.4$ is in Debye \cite{Bernath2005}. The dipole moment was calculated as \cite{vanhaecke2007precision}
\begin{equation}
\mathbf{M}_{200-100}=\sqrt{\frac{\hbar}{m_{{\rm red}}\omega_{1}}}\left[\frac{d\mathbf{M}_{200-100}}{dR}\right]_{R=R_{e}}
\end{equation}
where we used the approximate value for the slope of the dipole moment at the equilibrium separation of $3.17\,D/a_{0}$ estimated for the isoelectronic molecule SrF. The resulting lifetime is $1/A_{200-100} \approx$ 140 ms. The black body induced decay rate $\Gamma_{{\rm BBR}}$ is further suppressed by a factor $1/\left(\exp\left(\hbar\omega_{1}/\left(k_{B}T\right)\right)-1\right)\approx0.1$ at room temperature \cite{buhmann2008surface}.

%\section{Insensitivity to systematic errors}\label{sec:systematics}
\section{Estimation of systematic errors}\label{sec:systematics}

Here we show that several anticipated systematic errors can be suppressed to below the target measurement precision, owing to the large enhancement factors and favorable molecular structure of SrOH.

\subsection{Line broadening and shift}

As previously experimentally demonstrated with atomic microwave clocks \cite{bauch2003caesium} and theoretically analyzed for a YbF molecular fountain \cite{Tarbutt2013}, laser-cooled samples provide excellent suppression of possible systematic errors in precision measurement experiments. Doppler broadening is given by \cite{Bernath2005} 
\begin{equation}
\Delta\omega_{{\rm D}}=2\omega\sqrt{\frac{2kT\ln\left(2\right)}{m c^{2}}}
\end{equation}
and will be suppressed at ultracold temperatures ($\sim50\,{\rm \mu K}$) to $\triangle\omega_{{\rm D}}\approx5\times10^{-10}\omega$, which is 2 orders of magnitude lower than for a 1 K sample of SrOH and below the natural linewidth $1/T_{c}$ for $\omega_{200-03^{1}0}=2\pi\times1.2$ GHz, illustrating one advantage of driving a transition between near-degenerate states to suppress systematic effects. The second order relativistic Doppler shift is proportional to $v_{{\rm thermal}}^{2}/c^{2}$ and will be $\triangle\omega_{{\rm RD}}\approx10^{-20}\omega$ for an ultracold SrOH sample.

Blackbody radiation (BBR) can cause AC Stark shifts of molecular energy levels. In order to determine whether BBR-induced light shifts will cause an issue for the proposed measurements we need to consider the differential BBR-shift for the two ro-vibrational levels under consideration as well as the experimentally viable value for the time-stability of the black-body environment surrounding the molecular cloud. The frequency shift for each level under consideration is \cite{vanhaecke2007precision}

\begin{equation}
    \triangle^{\mathrm{BBR}}_i=\frac{4\pi}{3\epsilon_0 h c^3h}\sum_j P\int d\nu\frac{\nu^3}{\exp{\left(h\nu/k_B T\right)-1}}\frac{\mathbf{M}^2_{i j}}{\nu-\nu_{i j}}.
    \label{eq:BBR-shift}
\end{equation}

In order to estimate the magnitude of $\triangle^{\mathrm{BBR}}$, we can recast Eq. \ref{eq:BBR-shift} in terms of convenient experimental units,

\begin{equation}
    \triangle^{\mathrm{BBR}}_i = \frac{3.136\times 10^{-7}}{4\pi^2}\tilde{T}^3\sum_j\mathbf{M}^2_{ij}F\left(\frac{\tilde{\omega}_{ij}}{\tilde{T}}\right),
\end{equation}
where $\tilde{T}$ is the temperature in cm\textsuperscript{-1}, $\mathbf{M}$ is in Debye and $F\left(y\right)$ is an integral function introduced by Farley and Wing \cite{farley1981accurate} to evaluate the BBR-induced shift in the case of an $E1$ transition. Since the BBR spectrum peaks around 600 cm\textsuperscript{-1}, which happens to be close to vibrational transitions in SrOH, we consider BBR-induced shifts due to vibrational transition resonances. For room temperature, $\tilde{T}\approx 200$ cm\textsuperscript{-1}, and using $\tilde{\omega}_1\approx 530$ cm\textsuperscript{-1} and previously estimated $\mathbf{M}_{v''v'}\approx 0.4$ Debye (see Sec.~\ref{sec:lifetime-estimate}), we obtain $\triangle^{\mathrm{BBR}}\sim1\,$mHz. This is consistent with the estimations provided in Ref. \cite{vanhaecke2007precision} for other similar molecules. For the rotational transitions $\tilde{\omega}_{ij}\ll \tilde{T}$, we obtain an asymptotic expression $F\left(y\right)\simeq - \pi^2 y/3$ \cite{farley1981accurate} and $\triangle^{\mathrm{BBR}}_i\lesssim1\,$mHz. 

While the absolute magnitude of the BBR shifts seems to be large for a given ro-vibrational state, Vanhaecke and Dulieu pointed out that the differential dynamic BBR shift $\triangle^{\mathrm{BBR}}_i-\triangle^{\mathrm{BBR}}_j$ for molecular ro-vibrational transitions can be have a relative uncertainty of $\sim 10^{-13}-10^{-14}$ \cite{vanhaecke2007precision}. In particular, the vibrational dependence of the molecular dipole moments for simple polyatomic molecules is on the order of part per hundred \cite{sharipov2017influence} and therefore potentially leads to a contribution to the differential BBR shift at the level of $\sim10\,\mu$Hz, %\red{above it says 1 mHz; a factor of 100 down is 10 $\mu$ Hz}
 which is on the order of the absolute statistical uncertainty for one day of experimental integration. Atomic clock experiments have characterized the magnitude of BBR shifts with a fractional uncertainty of $\lesssim 5\times 10^{-4}$ \cite{mcgrew2018atomic}. We do not anticipate any significant BBR anisotropies (like in trapped molecular ion experiments, for example \cite{kokish2018prospects}). Assuming realistically that BBR shifts can be characterized at the part per thousand level, even in the worst case scenario of estimated absolute shift $\triangle^{\mathrm{BBR}}\sim 1$ mHz, the resulting fractional uncertainty in the transition frequency measurement will be $\delta\omega^{\mathrm{BBR}}/\omega\lesssim 10^{-15}$, thus not limiting the experimental precision at the level of $\delta\mu/\mu \sim 10^{-17}$ due to the large enhancement factor $Q_\mu\gg 100$. A recent work by Norrgard and co-workers describes a method to use molecules with optical cycling properties to perform quantum blackbody thermometry with temperature sensitivity of $\sigma_{T}/T\approx10^{-4}-10^{-5}$ \cite{norrgard2020quantum}. Using such methods to perform \textit{in situ} measurements of $T_{\mathrm{BBR}}$ with trapped SrOH would allow further control over the BBR-induced systematics and enable $\delta\triangle_i^{\mathrm{BBR}}\ll 1\mu$Hz.

\subsection{Field-insensitive transitions}

To assess the sensitivity of our proposed approach to electric- and magnetic-field-induced systematic errors, we compute the full energy level structure of the $\tilde{X}(200)$ and $\tilde{X}(03^10)$ states up through the lowest 3 rotational levels in each state ($N=2$ and $N=3$, respectively). The Hamiltonian takes the form
\begin{equation}
    H=H_\mathrm{rot}+H_\mathrm{SR}+H_\mathrm{\ell}+H_\mathrm{Fermi}+H_\mathrm{dd}+H_\mathrm{Stark}+H_\mathrm{Zeeman},
\end{equation}
where $H_{\mathrm{rot}}$ is the rotational Hamiltonian, $H_{\rm{SR}}$ is the spin-rotation Hamiltonian, $H_{\rm{\ell}}$ is the $\ell$-doubling Hamiltonian applicable to the $(03^10)$ state, $H_\mathrm{Fermi}$ is the Fermi contact hyperfine interaction, $H_{\rm{dd}}$ arises from the electron spin-nuclear spin dipole-dipole interaction, $H_\mathrm{Stark}$ is the Stark interaction, and $H_\mathrm{Zeeman}$ is the Zeeman interaction. Matrix elements for the rotational, spin-rotation, and $\ell$-doubling Hamiltonians may be found in \cite{merer1971rotational} for both the bending and non-bending vibrational states. The hyperfine Hamiltonians are $H_\mathrm{Fermi}=b_F I\cdot S$ and $H_\mathrm{dd}=c(I_zS_z-I\cdot S/3)$ \cite{fletcher1993molecular}, with matrix elements found in Ref. \cite{Hirota1985}. Likewise, the Stark interaction matrix elements may be found in Ref. \cite{Hirota1985}. Following~\cite{nguyen2018fluorescence}, we use $H_\mathrm{Zeeman}=g_S\mu_B S\cdot \mathcal{B} + g_l\mu_B(S_x\mathcal{B}_x+S_y\mathcal{B}_y)$ where $S_{x(y)}$  and $\mathcal{B}_{x(y)}$ are written in the molecular frame. The electron $g$-factor is constrained to its nominal value of $g_S=2.002319$, and $g_l=-\gamma/(2B)$ is given by the Curl identity. The relevant matrix elements are available in Ref. \cite{Hirota1985}, where we can use standard methods to transform $\mathcal{B}$ from the lab frame to the molecule frame \cite{BrownCarrington}.

While most of these Hamiltonians and matrix elements are readily available in the literature, it would be easy to overlook that the general form of the spin-rotation Hamiltonian is $H_\mathrm{SR}=\gamma\, S\cdot(J-S-G)$, where $\gamma$ is the spin-rotation constant, $J$ is the total angular momentum excluding nuclear spin and $G$ is the angular momentum associated with the vibrational motion \cite{merer1971rotational}. As a result, in the bending vibrational state, $H_\mathrm{SR}$ includes an interaction between the electron spin and vibrational angular momentum along the molecular symmetry axis. The spin-rotation splitting of a rotational state $N$ in $(03^10)$ is therefore given by
\begin{equation}
    \Delta E(N)=\frac{\gamma}{2}\left(2N+1-\frac{1}{N}-\frac{1}{N+1}\right).
\end{equation}
In a non-bending mode, the last two terms, which decrease with higher $N$, are absent. In addition, these terms are sometimes neglected for spectroscopy of bending modes when the high-$N$ limit is appropriate. However, for the low-$N$ states of interest here, all terms must be retained.

The rotational constant $B$, spin-rotation constant $\gamma$, $\ell$-doubling constant $q$, Fermi contact coefficient $b_F$, dipole-dipole hyperfine coefficient $c$, and electric dipole moment $D$ have been previously measured for SrOH and are given in Tab.~\ref{tab:molecular-cosntants} with appropriate references for both the (200) and $(03^10)$ states.

As an example of this system's robustness against systematic errors, we will first consider transitions between the $N''=1\rightarrow N'=1$ manifolds of the $\tilde{X}(200)\rightarrow\tilde{X}(03^10)$ transition.

In our numerical calculations, a reliable estimate of the energy shifts in state $N$ requires explicit diagonalization of the Hamiltonian including states up through $N+1$. This can be understood as follows. At vanishing electric field, $\mathcal{E}=0$, the dipole moment of every state is zero. In the (200) manifold, the mixing of nearby rotational states at non-zero electric field induces a dipole moment in each sublevel. As a result, the calculated induced dipole moments at low field are only valid for state $N$ if the basis includes the $N+1$ state. Although the induced dipole moments in $(03^10)$ states are dominated by mixing within $\ell$-doublets, rotational mixing occurs at the same level as for the $(200)$ manifold and therefore must also be considered to obtain an accurate estimate of the electric field sensitivity of a transition between the $(200)$ and $(03^10)$ manifolds. In the calculations presented here, our basis spans up to $N=2$ in $(200)$ and $N=3$ in $(03^10)$, for a total of 156 states.

%\begin{widetext}
\begin{table}
\begin{tabular}{@{}lll@{}}
\toprule
\textbf{Parameter} & $\mathbf{(200)}$               & $\mathbf{(03^10)}$ \\ \midrule
$B$ [MHz]       &  7,384.788 \cite{presunka1995laser}              & 7,429.631 \cite{fletcher1995excited}          \\
$\gamma$ [MHz]  & 72.77 \cite{presunka1995laser} & 71.14 \cite{fletcher1995excited}          \\
$q$ [MHz]      & 0                      & $-12.484$ \cite{fletcher1995excited} \\
$b$ [MHz]      & 1.713 \cite{fletcher1993molecular}                      & 1.713  \cite{fletcher1993molecular}           \\
$c$ [MHz]       & 1.673 \cite{fletcher1993molecular}                      & 1.673 \cite{fletcher1993molecular}          \\
$D$ [Debye]       & 1.900 \cite{steimle1992supersonic}                      & 1.900 \cite{steimle1992supersonic}\\ \bottomrule     
\end{tabular}
\caption{\label{tab:molecular-cosntants}Molecular constants used to calculate electric and magnetic transition sensitivity}
\end{table}
%\end{widetext}

The measurement will be robust against systematic errors related to electric (magnetic) field drifts if the transition has a negligible difference between the ground and excited state electric (magnetic) dipole moments. We numerically diagonalize the full Hamiltonian in each vibrational manifold at a variety of magnetic and electric fields and compute the local dipole moments of each sublevel from the change in energy with respect to field strength. In considering the relative dipole moments between two states, we restrict our attention to those whose overall transition strength is at least a non-negligible fraction (e.g., $10$\%) of the strongest transition.

\begin{figure*}
\begin{centering}
\includegraphics[width=8.7cm]{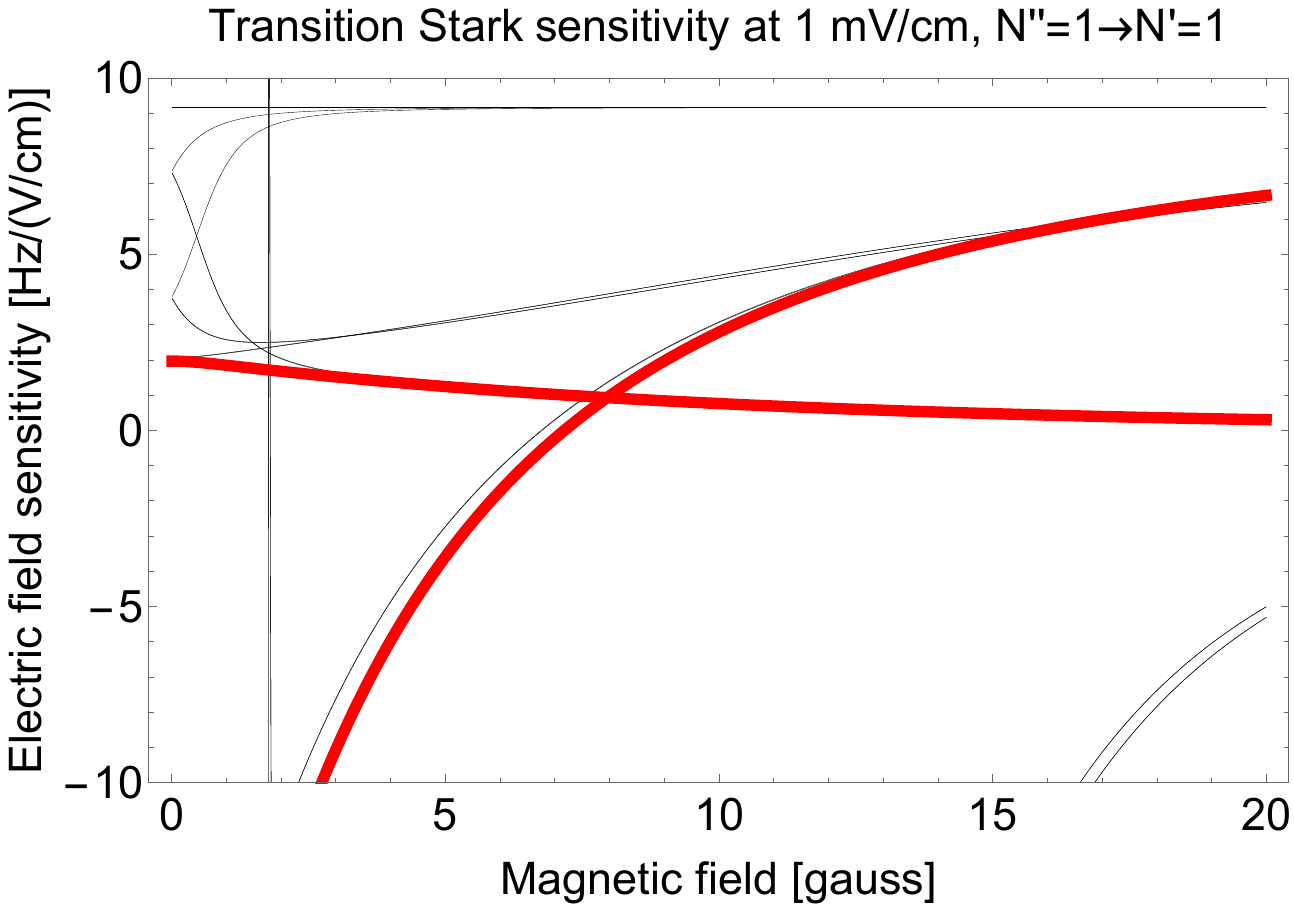} \includegraphics[width=8.7cm]{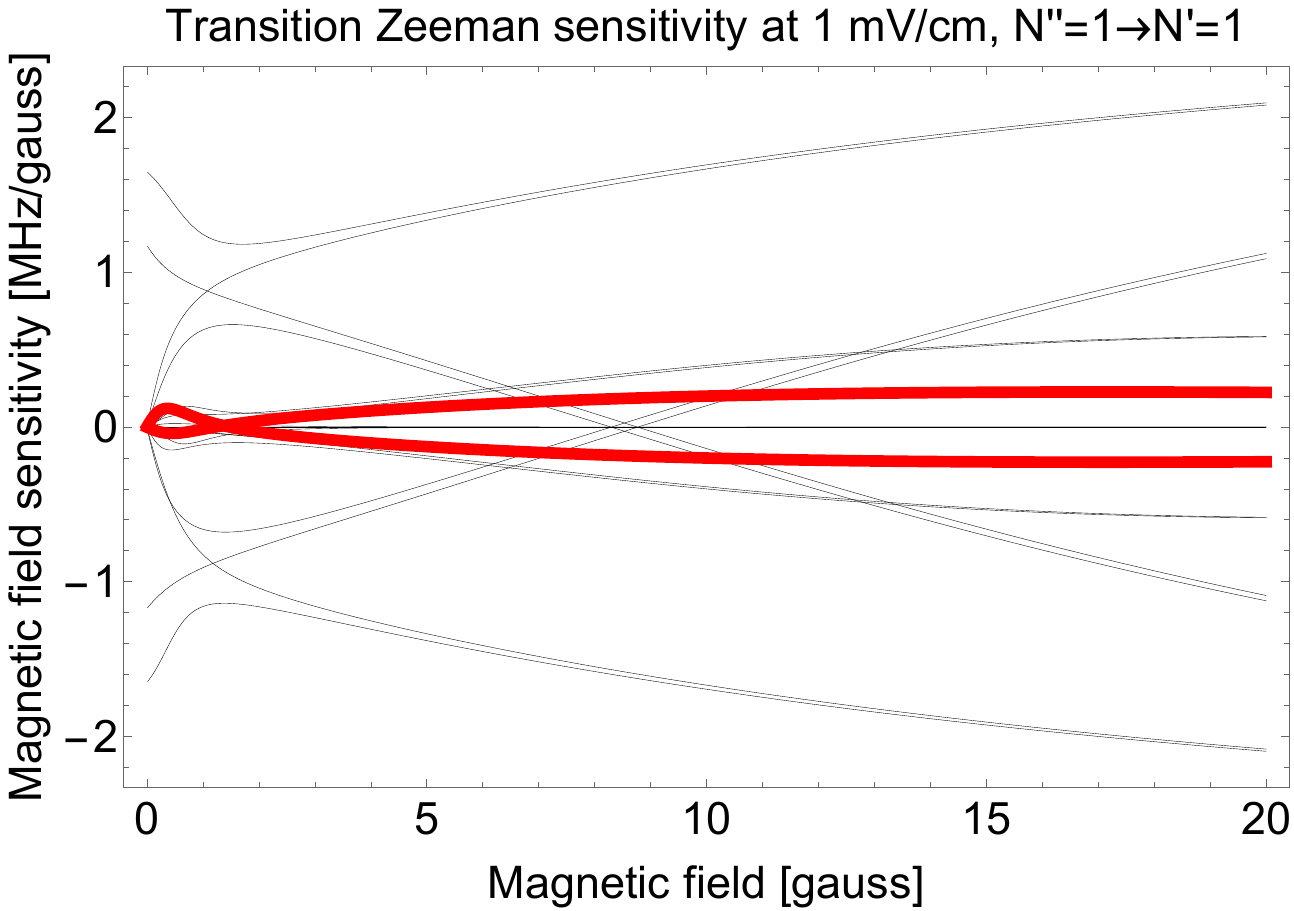}
\par\end{centering}
\caption{\label{fig:stark11} Relative electric (left) and magnetic (right) dipole moments of strong transitions among $N''=1\rightarrow N'=1$. The thick, red lines show two transitions with low relative electric dipole moments and nearly zero common mode sensitivity to electric or magnetic fields around 6.40 G.}
\end{figure*}

% \begin{figure}
% \begin{centering}
% \includegraphics[width=8cm]{11Stark.pdf} 
% \par\end{centering}
% \caption{\label{fig:stark11}Relative electric dipole moments of strong transitions among $N''=1\rightarrow N'=1$. The thick, red lines show two transitions with low relative electric dipole moments and nearly zero common mode sensitivity to electric or magnetic fields around 6.40 G.}
% \end{figure}

% \begin{figure}
% \begin{centering}
% \includegraphics[width=8cm]{11Zeeman.pdf} 
% \par\end{centering}
% \caption{\label{fig:zeeman11}Relative magnetic dipole moments of strong transitions among $N''=1\rightarrow N'=1$. The thick, red lines show two transitions with low relative magnetic dipole moments and nearly zero common mode sensitivity to electric or magnetic fields around 6.40 G.}
% \end{figure}

See Fig.~\ref{fig:stark11} for the relative dipole moments of strong transitions among $N''=1\rightarrow N'=1$. The sharp vertical line in Fig.~\ref{fig:stark11} arises from a resonance between opposite-parity states in $(03^10)$ as the magnetic field is tuned. The thick, red transitions have the approximate composition
\begin{multline}
    |(200),N''=1,J''=1/2,F''=0-1,M''=0\rangle \\
    \rightarrow |(03^10),N'=1,J'=1/2,F'=0-1,M'=0\rangle
\end{multline}
and
\begin{multline}
    |(200),N''=1,J''=3/2,F''=1-2,M''=0\rangle \\
    \rightarrow |(03^10),N'=1,J'=3/2,F'=1-2,M'=0\rangle,
\end{multline}
where a dash denotes an even superposition of different $F$ states.

In addition to having comparatively small individual relative dipole moments, these highlighted transitions can be made to have nearly exactly opposite sensitivities to both electric and magnetic fields. In particular, they have relative $g$-factors of $+0.1105$ and $-0.1099$ at fields of 6.40 G and 1 mV/cm, for a common-mode sensitivity to magnetic fields characterized by $\Delta g/2\sim3\times10^{-4}$. Therefore, simultaneously measuring the resonance frequency of both transitions, and averaging the results, allows near complete elimination of magnetic field-induced systematic errors. Although many pairs of opposite-magnetic-sensitivity transitions exist, it is typically the case that such pairs of transitions have large individual and common-mode electric field sensitivity at any particular magnetic field; thus simultaneously suppressed common-mode electric and magnetic relative dipole moments are non-trivial and must be found numerically.
The above-estimated common-mode magnetic dipole moment is smaller than the uncertainty arising from existing Zeeman spectroscopy of SrOH. In particular, the rotational and nuclear $g$-factors, expected to be of order $10^{-3}$, have not yet been measured in SrOH. With refined measurements of the Zeeman structure, the optimal conditions to minimize the common-mode sensitivity to magnetic fields could be fine-tuned.

In a similar manner, the two transitions considered above have nearly opposite electric polarizabilities of $-1081$ Hz/(V/cm)$^2$ and $+1085$ Hz/(V/cm)$^2$, for an average polarizability of only $2\,\mu$Hz/(mV/cm)$^2$ at a magnetic field of 6.40 G. A transition in SrOH, or other co-trapped species, with several hundred times greater sensitivity to electric fields could be used as a reference to actively stabilize the electric field over the small volume of the optical dipole trap to the mV/cm level, thus reducing systematic errors in common-mode resonance to the $\mu$Hz level, which is below the frequency uncertainty obtainable with one day of experimental integration. The electric dipole moment can be fine-tuned, and its sign can be reversed, with changes in magnetic field on the order of 1$-$10 mG.

\begin{figure*}
\begin{centering}
\includegraphics[width=8.7cm]{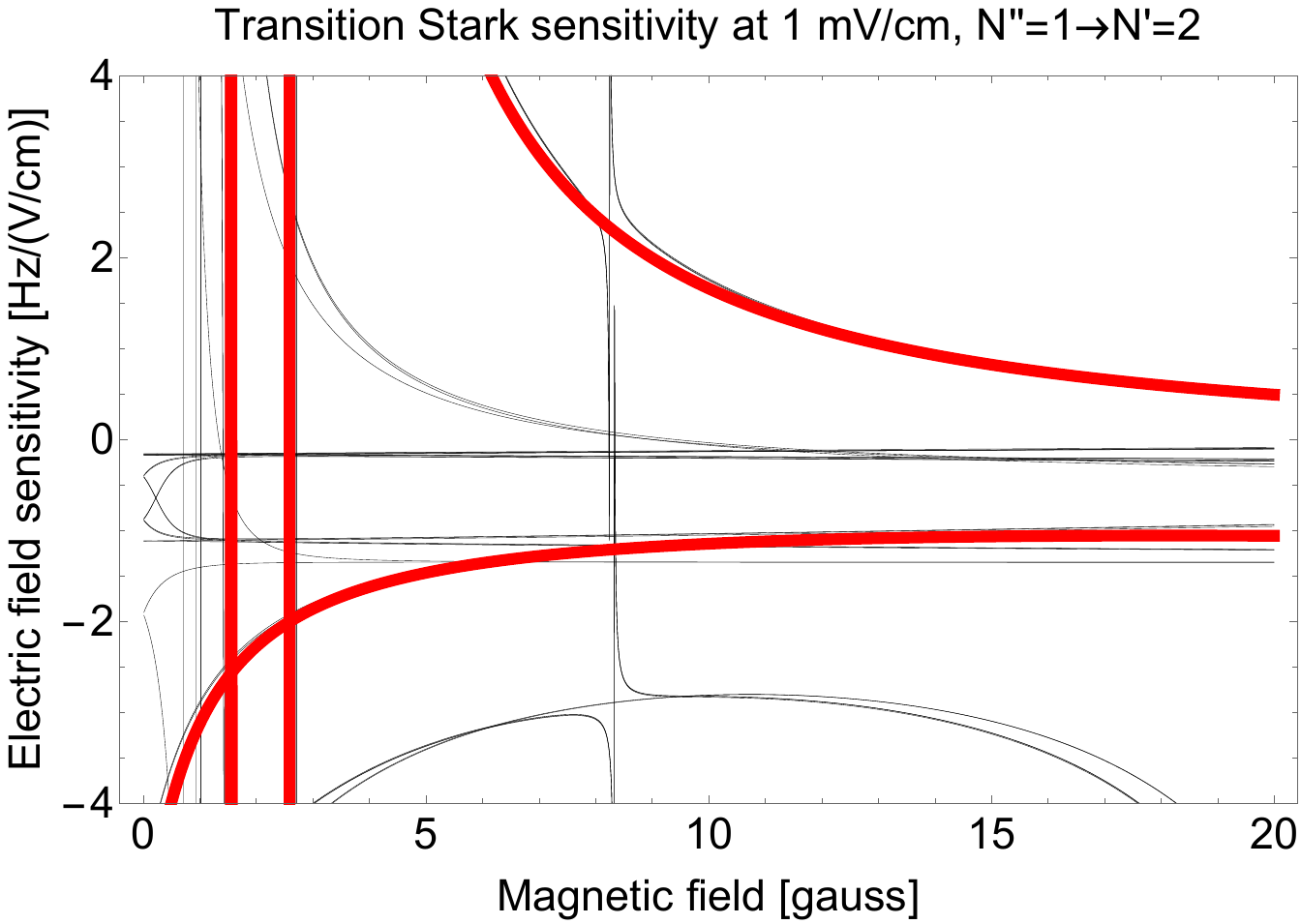} \includegraphics[width=8.7cm]{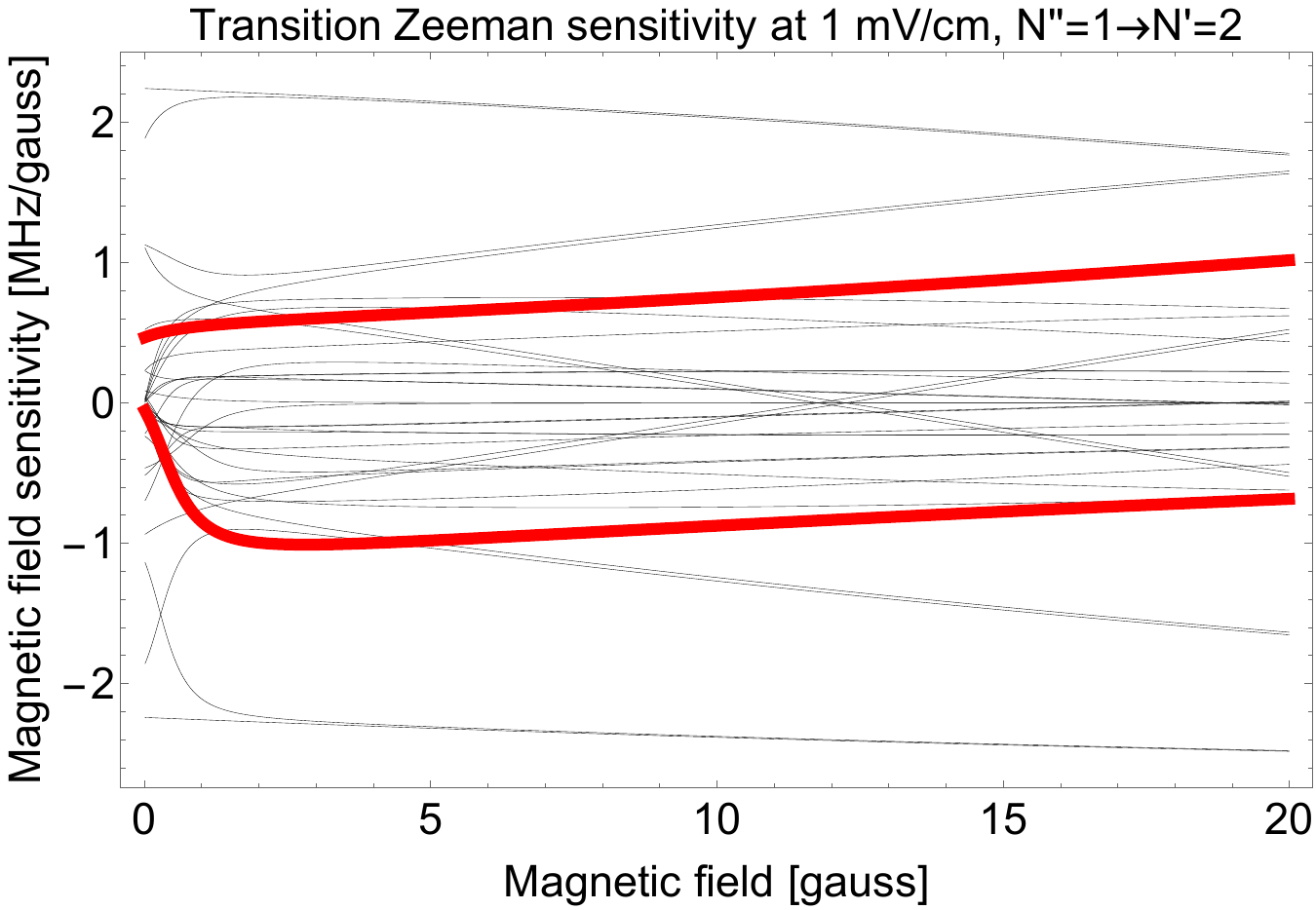}
\par\end{centering}
\caption{\label{fig:stark12} Relative electric (left) and magnetic (right) dipole moments of strong transitions among $N''=1\rightarrow N'=2$. The thick, red lines show two transitions with low relative electric dipole moments and nearly zero common mode sensitivity to electric or magnetic fields around 12.75 G.}
\end{figure*}

% \begin{figure}
% \begin{centering}
% \includegraphics[width=8cm]{12Stark.pdf} 
% \par\end{centering}
% \caption{\label{fig:stark12}Relative electric dipole moments of strong transitions among $N''=1\rightarrow N'=2$. The thick, red lines show two transitions with low relative electric dipole moments and nearly zero common mode sensitivity to electric or magnetic fields around 12.75 G.}
% \end{figure}

% \begin{figure}
% \begin{centering}
% \includegraphics[width=8cm]{12Zeeman.pdf} 
% \par\end{centering}
% \caption{\label{fig:zeeman12}Relative magnetic dipole moments of strong transitions among $N''=1\rightarrow N'=2$. The thick, red lines show two transitions with low relative magnetic dipole moments and nearly zero common mode sensitivity to electric or magnetic fields around 12.75 G.}
% \end{figure}

We have identified additional favorable transitions in the range of 0$-$20 G at fields around 5.94 G, 18.75 G, and 19.07 G for the $N''=1\rightarrow N'=1$ manifold. Furthermore, it is straightforward to find transitions between other rotational manifolds with suppressed sensitivity to systematic errors. As an example, see Fig.~\ref{fig:stark12} for a pair of transitions in the $N''=1\rightarrow N'=2$ manifold with a nominal average g-factor of $\Delta g/2\sim4\times 10^{-4}$ and average polarizability of $-6\,\mu$Hz/(mV/cm)$^2$. Once again, this magnetic moment is consistent with 0 at the level of existing spectroscopy and the average electric sensitivity of these transitions can be fine-tuned and reversed with small adjustments of the magnetic field. Comparably favorable transitions have been found for the $N''=0\rightarrow N'=1$ rotational transition.

%\subsection{BBR-induced shifts}

\section{Sensing cosmic fields}\label{sec:cosmic-fields}

%additional aspects of the SrOH spectrum could be used to probe other
%important $m_{\phi}$ ranges. %Resonant absorption of bosonic dark
%matter in atoms and molecules has recently been explored theoretically
%\cite{sikivie2014axion}. With long coherence times and different
%energy level spacings in the megahertz, gigahertz and terahertz frequency
%range, trapped ultracold SrOH provides a potentially promising platform
%for exploring coherent internal transitions induced by dark matter
%fields (SM). 

% \subsection{Sensitivity to new physics}

The proposed measurement is predominantly sensitive to oscillation frequencies between the inverse of the total measurement time (e.g., 1 day or 1 year) and the $\tilde{X}(200)$ decay rate. We perform least-squares spectral analysis (LSSA) on simulated data sets to quantify the projected sensitivity~\cite{Cumming1999lick,Cumming2004detectability,abel2017search}. This method is closely related to the discrete Fourier transform but can be applied to the experimentally realistic situation in which data are not uniformly distributed in time, and accommodates inspection of arbitrary oscillation frequencies. We briefly summarize the LSSA approach here. For a discrete series of measurements, {$\omega(t_i)$}, made at times $\{t_i\}$, we fit the data to a model $\omega(t_i)=A_j\sin(2\pi f_j t_i)+B_j\cos(2\pi f_j t)+C_j$, where $A_j,\,B_j,$ and $C_j$ are fit parameters and $f_j$ is a possible oscillation frequency of the resonance. The estimated amplitude of oscillation at frequency $f_j$ is then $\widehat{\delta\omega}(f_j;\delta\omega)=\sqrt{A_j^2+B_j^2}$, where $\delta\omega$ is the true oscillation amplitude at $f_j$. This procedure is repeated for each oscillation frequency $f_j$ that is of interest.

For our simulation, we suppose that $N=10^6$ trapped molecules are probed approximately every coherence time $T_{c}=140$ ms, with random delays of order $\sim0.1 T_c$ between subsequent measurements. The single-measurement frequency sensitivity is assumed to be shot-noise limited, with statistical uncertainty $\Delta\omega = (T_c \sqrt{N})^{-1}$~\cite{hanneke2016high,ludlow2015optical}. We first simulate the case of a single series of measurements, $\omega(t_i)$, over 24 hours, with no assumed oscillation of the resonance frequency. The inferred values of $\widehat{\delta\omega}(f_j;\mathrm{0})>0$ arise from statistical noise and allow an estimation of the noise floor of the measurement, $\Delta\delta\omega(f_j)$. In the case of measurements over one day, we find a noise floor of $\Delta\delta\omega(f_j)\approx 2\pi\times3$ $\mu$Hz for $f\gtrsim10$ $\mu$Hz, as expected from the shot-noise limit. At low frequencies, $f_j\lesssim 10$ $\mu$Hz, the sensitivity falls off as $f_j^{-2}$ because the inverse of the total measurement time, 24 hours, is $\sim10$ $\mu$Hz and lower frequencies cannot be resolved from an offset in the mean of the resonance.

To calculate the sensitivity to oscillation of the resonance at frequency $f_j$, we then simulate a series of measurements with a large oscillating resonance $\omega(t)=\omega_0+\delta\omega(f_j)\sin(2\pi f_j t+\phi_j)$, where $\phi_j\in [0,2\pi)$ is chosen randomly and we set $\delta\omega(f_j)=2\pi\times 1$ Hz. Each measurement of $N$ molecules gives a measured value of $\omega$ equal to the mean of $\omega(t)$ over the 140 ms duration of the measurement, up to statistical shot noise. At all sufficiently low oscillation frequencies, $f_j\lesssim (2\pi T_c)^{-1}\sim1$ Hz, the inferred amplitude of oscillation is accurate to excellent precision, $\widehat{\delta\omega}(f_j;\mathrm{1\:Hz})\approx1$ Hz. At large oscillation frequencies, $f_j\gtrsim(2\pi T_c)^{-1}$, the sensitivity falls off approximately as $\widehat{\delta\omega}(f_j;\mathrm{1\:Hz})\propto f_j^{-1.4}$ because the average shift in the resonance frequency averages toward 0 over many oscillations.

The oscillation amplitude $\delta\omega_{\mathrm{SNR}=1}(f_j)$ at frequency $f_j$ that would generate a measurement signal-to-noise ratio (SNR) of 1 is then given by 
\begin{equation}
\frac{\delta\omega_{\mathrm{SNR}=1}(f_j)}{\mathrm{1\:Hz}}=\frac{\widehat{\delta\omega}(f_j;0)}{\widehat{\delta\omega}(f_j;\mathrm{1\:Hz})}.
\end{equation}

We repeat this procedure for the case of data interspersed throughout one year, with one 24-hour series of measurements repeated weekly. In this case, the sensitivity at intermediate and high frequencies improves by approximately $\sqrt{52}$ due to the shot-noise limit, and the low-frequency noise cutoff is reduced to $\sim$30 nHz, set by the inverse of the total measurement time of 1 year.

Using the estimated enhancement factor $Q_\mu\approx-617$ and transition frequency of 1.1 GHz, we find the fractional $\mu$-variation at frequency $f_j$ corresponding to a signal-to-noise ratio of unity, $(\delta\mu/\mu)_{\rm{SNR}=1}(f_j)=\delta\omega_{\mathrm{SNR}=1}(f_j)/(\omega Q_\mu)$. The oscillation frequency $f$ is related to the mass $m$ of the new scalar particle by $f=2.42\times10^5(m c^2/\rm{neV})$ Hz~\cite{derevianko2018network}.

The discussion above allows us to interpret the sensitivity of the measurement in terms of $\delta\mu/\mu$ as a function of the mass $m_\phi$ of a possible scalar dark matter particle $\phi$. To go further we must consider concrete models. As an example, we consider models of ultralight scalar particles with dilatonic interactions, characterized by coupling constants $d_{m_e}$, $d_g$, and $d_{\hat{m}}$, which arise from couplings of $\phi$ to electrons, gluons, and the symmetric combination of up and down quarks, respectively~\cite{arvanitaki2015searching}. Assuming the new scalar particle comprises all of dark matter,~\cite{arvanitaki2015searching,VanTilburn2016Thesis}

\begin{equation}\label{eq:dilatonic-coupling-sensitivity}
    \frac{\delta\mu}{\mu}=(d_{m_e}-d_g+M_Ad_{\hat{m}})\kappa\phi(t),
\end{equation}
where $\kappa\phi_0 = 6.4\times10^{-13}(10^{-18}\mathrm{\,eV}/m_\phi)$ with $\phi_0$ being the amplitude of the time-dependent dark matter field $\phi\left(t\right)$ and $M_A=0.037$ quantifies the variation of the nucleon mass with the quark mass in the case considered here of a transition directly sensitive to the proton-to-electron mass ratio \cite{flambaum2004limits}.

\begin{figure}
\begin{centering}
\includegraphics[width=8cm]{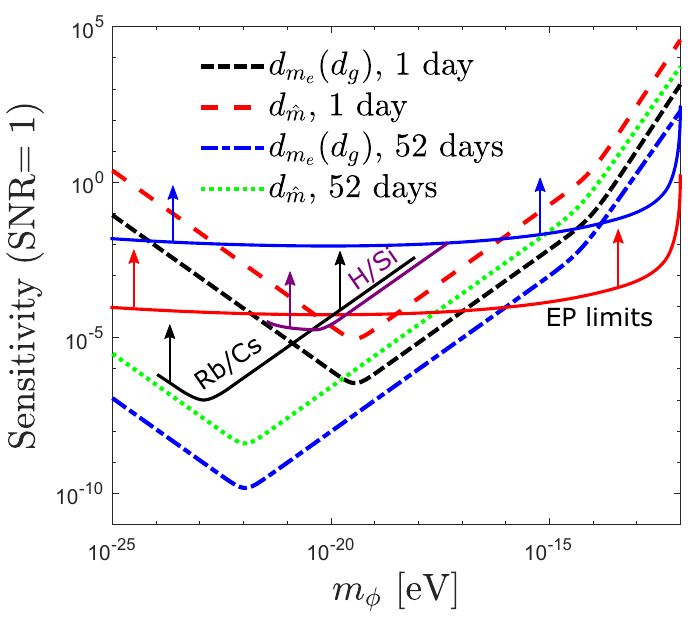} 
\par\end{centering}
\caption{\label{fig:sensitivity-plot} Dashed lines depict sensitivity of the proposed measurements to select ultralight dark matter couplings, assuming single one-day measurement, or 52 one-day measurements interspersed weekly over a year. The low-frequency cusp corresponds to masses with corresponding compton frequency at one day (year), below which sensitivity falls off rapidly, while the high-mass cusp corresponds to the decay rate of $\tilde{X}(200)$. Solid blue (upper) and red (lower) lines indicate existing limits from the equivalence principle (EP) tests for $d_{m_e}$ and $d_{\hat{m}}$ terms, respectively \cite{van2015search}. Solid black line shows limits on the $d_{\hat{m}}-d_g$ coupling term from dual Rb/Cs cold atomic fountain at LNE-SYRTE \cite{hees2016searching,hees2018violation}, while solid magenta line depicts a limit on $d_{m_e}$ from a comparison of a hydrogen maser with a crystalline silicone optical cavity \cite{kennedy2020precision}.}
\end{figure}

From Eq.~\ref{eq:dilatonic-coupling-sensitivity} we can interpret the experimental sensitivity to $\mu$ variation in terms of sensitivity to $d_{m_e}$, $d_g$, and $d_{\hat{m}}$. Because $|M_A|<1$, the parameter space probed for $d_{\hat{m}}$ is less stringent than for $d_{m_e}$ and $d_g$. The sensitivity to these parameters is shown as a function of $m_\phi$ in Fig.~\ref{fig:sensitivity-plot}. For comparison, we also show existing bounds on $d_{m_e}$ and $d_{\hat{m}}$ obtained from equivalence principle (EP) tests.

The proposed measurement with SrOH would improve on the EP tests by up to about 8 orders of magnitude at the most sensitive frequency with 52 days of data, or over almost a decade in mass range with only 24 hours of data. The largest sensitivity to the coupling coefficient $\Gamma_{\mu}$ between potential UDM coherent oscillations and proton-to-electron mass ratio in a one-day measurement occurs for dark matter particles in the mass range $m_{\phi}\sim5\times10^{-20}\,{\rm eV}$ to $1\times10^{-14}\,{\rm eV}$, corresponding to oscillation periods of one day and the Nyquist period $2T_{c}$, respectively. If measurements are interspersed throughout a year, masses as low as $1\times10^{-22}\,{\rm eV}$ can be probed \cite{derevianko2018network}. These mass ranges and the coupling coefficients in the range shown in Fig.~\ref{fig:sensitivity-plot} are already of interest to fundamental particle physics \cite{fuzzyDM,marsh2015axion,lora2012mass}. The use of quantum enhanced metrology methods experimentally demonstrated for microwave clocks can lead to further gains in sensitivity \cite{kasevich2016measurement}. For $m_{\phi}\gtrsim10^{-23}\,$eV, our projected limits for 52 days of integration will improve existing experimental bounds from atomic spectroscopy over a 6-year time period \cite{hees2016searching, hees2018violation} by 4 orders of magnitude and will be complementary to future proposed searches using atomic clocks as they probe different combinations of the coupling constants $d_i$  \cite{arvanitaki2015searching}.

% Figure \ref{fig:sensitivity-plot} summarizes the expected sensitivity of the proposed experiment to various dimensionless coefficients that characterize the coupling between the scalar cosmic fields and standard model particles $\Gamma_{\mu}\simeq \left(d_{m_e}-d_g+M_Ad_{\hat{m}}\right)\sqrt{4\pi}/M_{\rm{Pl}}$ with $d_{m_e}$, $d_g$ and $d_{\hat{m}}$ quantifying coupling to the electron mass, gluon field, and the symmetric combination of the light-quark masses \cite{arvanitaki2015searching,flambaum2004limits}.

\section{Summary}

We have considered the search for ultralight dark matter using precision microwave spectroscopy of the laser-cooled triatomic radical SrOH. The rovibronic spectrum of SrOH in the ground electronic state has been analyzed, and the enhancement factors $Q_{\mu}$ are calculated for different rotational transitions in the $\left(200\right)\leftrightarrow\left(03^{1}0\right)$ vibrational band. With predicted $\left|Q_{\mu}\right|\gg10$ for multiple rovibrational transitions, as well as highly diagonal Franck-Condon factors in the $\tilde{X}\leftrightarrow\tilde{A}$ electronic excitation band, laser-cooled SrOH provides a viable molecular platform for achieving $\sim10^{-17}$ uncertainty in $\delta\mu/\mu$ with 1 day of integration and has the potential to significantly improve on the previous limit on $\delta\mu/\mu$ from molecular spectroscopy \cite{shelkovnikov2008stability}.
%While microwave spectroscopy of low-lying excited vibrational modes
%of SrOH appears promising as a new experimental platform for UDM searches,
%accurate\textit{ ab initio} calculations and preliminary spectroscopy
%experiments are required to determine the exact transition frequencies
%as well as enhancement factors $Q_{\mu}$ estimated here.

%Other linear triatomic molecules with potentially diagonal Franck-Condon
%factors like RaOH and YbOH have been recently proposed for exploring
%fundamental physics of parity violation and permanent electric dipole
%moments \cite{isaev2017laser,kozyryev2017precision}. However, advanced
%molecular spectroscopy is still necessary for both species to make
%their laser cooling prospects practical and identify the optimal repumping
%schemes. Additionally, the presence of a heavy atom like Ra or Yb
%could lead to additional complications in the laser cooling and trapping
%schemes, similar to YbF \cite{lim2018laser} compared to SrF \cite{Shuman2010}.

Looking for signatures of high-energy physics in low-energy spectroscopy experiments with laser-cooled SrOH has the potential to complement other experimental efforts to uncover the quantum mechanical nature of the dark sector of the universe \cite{budker2017CASPEr,van2015search}. Furthermore, while SrOH is one of the simplest examples of monovalent metal alkoxides (MOR) that have been previously identified as suitable for direct laser cooling and trapping \cite{kozyryev2016MOR}, degeneracies between vibrational states of different character are ubiquitous among polyatomic molecules. For example, CaOH is another triatomic molecule which has since been laser cooled in a cryogenic beam and is actively being pursued for three-dimensional magneto-optical trapping~\cite{Baum2020magneto,Baum2020establishing}. Previous high-resolution vibrational spectroscopy~\cite{Coxon1992investigation,Li1995high,Baum2020establishing} of CaOH  predicts a transition energy of only $\sim0.1\,\rm{cm}^{-1}$ for the $Q(N=4)$ transition of $\tilde{X}(06^00)\rightarrow\tilde{X}(14^40)$, with an associated enhancement factor estimated to be $|Q_\mu|\sim500$. These states are subject to significant anharmonic contributions and Coriolis resonances, and differ by only 5 vibrational quanta like the states of interest for SrOH; it is therefore reasonable to expect a similarly strong transition moment as analyzed above. Thus further spectroscopy and characterization of these states may reveal an alternative route to probe $\delta\mu/\mu$ via precision measurement of rovibrational transitions in triatomic MOH molecules. 

The higher density of rovibrational states provided by the mechanical motion of MOR molecules with more complex ligands could result in similar degeneracies as analyzed here but with even larger enhancement factors $Q_{\mu}$, enabling access to a new UDM-coupling range by probing $\delta\mu/\mu$ fractional uncertainty in the $\lesssim10^{-18}$ regime. For example, recently laser cooled MOR-type symmetric top molecule calcium monomethoxide CaOCH\textsubscript{3} \cite{Mitra2020direct} possesses two nearly degenerate vibrational modes arising from the mechanical motion of the CH\textsubscript{3} group ($\omega_{\rm{vib}}\sim1,450$ cm\textsuperscript{-1}). Previous \textit{ab initio} calculations predict that the CH\textsubscript{3} umbrella ($a_1$ symmetry) and scissoring ($e$ symmetry) motions are less than 20 cm\textsuperscript{-1} apart \cite{kozyryevCaOCH3,jinjun2019CaOCH3}, which can be further reduced to $\sim1$ cm\textsuperscript{-1} by driving perpendicular rovibrational transitions with $K''=1\rightarrow K'=2$. While further experimental measurements are needed to identify the contributions from anharmonic parts of the potential in order to accurately predict the enhancement factors $Q_{\mu}$, the presence of new rotational degrees of freedom compared to linear molecules enables precise ``tuning'' of the separation between near-degenerate levels.   
%extending the maximum $m_{\phi}$ sensitivity. 

\begin{acknowledgments}
This work has been funded by the AFOSR Grant No. FA9550-15-1-0446 and NSF Grant No. PHY-1505961. We would like to thank M. Safronova for encouraging us to pursue the topic of dark matter effects in the spectra of laser-cooled polyatomic molecules and for a critical reading of the initial version of the manuscript. We would also like to acknowledge insightful discussions with J. Weinstein during the early stages of this work and thank B. Augenbraun for bringing to our attention the feasibility of ro-vibronic near-degeneracies in CaOH. We are grateful to J. K\l{}os and S. Kotochigova for sharing the vibrational potential of SrOH.

I.K. and Z.L. contributed equally to this work.
\end{acknowledgments}

\begin{appendix}

\section{Estimation of SrOH molecular constants}\label{sec:molecular-constants}

In the past, extensive molecular spectroscopy has been performed on SrOH with many vibrational and rotational parameters precisely measured \cite{presunka1995laser}. In a ground electronic state, vibrational energy levels of a linear triatomic molecule like SrOH are given by \cite{Bernath2005} 
\begin{widetext}
\begin{equation}
G\left(v_{1}v_{2}^{l}v_{3}\right)=\sum_{i=1}^{3}\omega_{i}\left(v_{i}+\frac{d_{i}}{2}\right)+\sum_{i=1}^{3}\sum_{k\geq i}x_{ik}\left(v_{i}+\frac{d_{i}}{2}\right)\left(v_{k}+\frac{d_{k}}{2}\right)+g_{22}l^{2}\label{eq:Vibrational_term_value}
\end{equation}
\end{widetext}
where $d_{i}=1$ for non-degenerate stretching vibrations ($v_{1}$ and $v_{3}$) and $d_{i}=2$ for the doubly-degenerate bending mode $v_{2}.$ For SrOH and other similar molecules, the low-lying vibrational motions are mostly harmonic and therefore $\omega_{i}\gg x_{ii},\,x_{ik}$ for $i\neq k$. Therefore, SrOH vibrational levels of experimental relevance are approximated by the following expression:

\begin{equation}
E_{v_{1}v_{2}v_{3}}\simeq\omega_{1}\left(v_{1}+\frac{1}{2}\right)+\omega_{2}\left(v_{2}+1\right)+\omega_{3}\left(v_{3}+\frac{1}{2}\right)\label{eq:triatomic_vib}
\end{equation}

\[
+x_{11}\left(v_{1}+\frac{1}{2}\right)^{2}+x_{22}\left(v_{2}+1\right)^{2}+x_{33}\left(v_{3}+\frac{1}{2}\right)^{2}+g_{22}l^{2}.
\]
Using Eq. \ref{eq:triatomic_vib} as well as the measured energies of the $\left(100\right)$, $\left(200\right)$, $\left(01^{1}0\right)$, $\left(02^{0}0\right)$ and $\left(02^{2}0\right)$ states \cite{presunka1995laser}, we can estimate all of the necessary harmonic ($\omega_{1}$ and $\omega_{2}$) as well as anharmonic ($x_{11}$, $x_{22}$ and $g_{22}$) constants. It is computationally convenient to reference all of the excited vibrational levels relative to the ground vibrational level 
\begin{equation}
E_{000}=\frac{\omega_{1}}{2}+\omega_{2}+\frac{\omega_{3}}{2}+\frac{x_{11}}{4}+x_{22}+\frac{x_{33}}{4}.
\end{equation}
The estimated vibrational constants (in cm\textsuperscript{-1}) are $\omega_{1}=531.900$, $x_{11}=-2.455$, $\omega_{2}=369.584$, $x_{22}=-4.485$ and $g_{22}=7.558$. With these extracted constants and using Eq. \ref{eq:triatomic_vib} for vibrational levels of SrOH, we predict positions of $E_{100}$, $E_{200}$, $E_{01^{1}0}$, $E_{02^{0}0}$ and $E_{02^{2}0}$ to 0.002 cm\textsuperscript{-1}, which corresponds to 0.06 GHz. In particular, we have the following expressions (in units of cm\textsuperscript{-1}):
\begin{equation}
E_{100-000}=\omega_{1}+2x_{11}=526.991
\end{equation}
\begin{equation}
\triangle E_{200-100}=\omega_{1}+4x_{11}=522.082
\end{equation}
\begin{equation}
\triangle E_{02^{2}0-02^{0}0}=4g_{22}=30.233
\end{equation}
\begin{equation}
E_{01^{1}0-000}=\omega_{2}+3x_{22}+g_{22}=363.687
\end{equation}
\begin{equation}
E_{02^{0}0-000}=2\omega_{2}+8x_{22}=703.288.
\end{equation}

\section{Normal modes of a linear triatomic molecule}

In order to determine the dependence of vibrational frequencies of SrOH on the proton-to-electron mass ratio $\mu$, we perform the normal mode analysis using the $\mathbf{GF}$ matrix formalism \cite{wilson1955molecular}. The kinetic-energy-related matrix $\mathbb{\mathbf{G}}$ for a linear triatomic molecule is given by 
\begin{equation}
\mathbf{G}=\left[\begin{array}{ccc}
\mu_{1}+\mu_{2} & -\mu_{3} & 0\\
-\mu_{3} & \mu_{2}+\mu_{2} & 0\\
0 & 0 & G_{33}
\end{array}\right],
\end{equation}
where following a common convention in the literature we use the notation $\mu_{1}\equiv1/m_{{\rm Sr}}$, $\mu_{2}=1/m_{{\rm H}}$ and $\mu_{3}\equiv1/m_{{\rm O}}$ while 
\begin{equation}
G_{33}=\mu_{1}\frac{r_{32}}{r_{31}}+\mu_{2}\frac{r_{31}}{r_{32}}+\mu_{3}\frac{\left(r_{31}+r_{32}\right)^{2}}{r_{31}r_{32}},
\end{equation}
which also has units of 1/{[}mass{]}. The diagonal force constant matrix is given by 
\begin{equation}
\mathbf{F}=\left[\begin{array}{ccc}
F_{11} & 0 & 0\\
0 & F_{22} & 0\\
0 & 0 & F_{33}
\end{array}\right].\label{eq:Force_matrix}
\end{equation}
Solving for eigenvalues of $\mathbf{GF}$ and setting them equal to $\omega_{i}^{2}$, we can find an expression for the harmonic vibrational frequencies in terms of atomic masses: 
\begin{widetext}
\begin{eqnarray*}
\omega_{1,3}^{2} & =\left\{ F_{11}\left(\mu_{1}+\mu_{3}\right)+F_{22}\left(\mu_{2}+\mu_{3}\right)\mp\left(F_{11}^{2}\left(\mu_{1}+\mu_{3}\right)^{2}+F_{22}^{2}\left(\mu_{2}+\mu_{3}\right)^{2}+4F_{11}F_{22}\mu_{33}^{2}-2F_{11}F_{22}\left(\mu_{1}+\mu_{3}\right)\left(\mu_{2}+\mu_{3}\right)\right)^{\frac{1}{2}}\right\} 
\end{eqnarray*}
\end{widetext}
\begin{equation}
\omega_{2}^{2}=F_{33}\left(\mu_{1}\frac{r_{32}}{r_{31}}+\mu_{2}\frac{r_{31}}{r_{32}}+\mu_{3}\frac{\left(r_{31}+r_{32}\right)^{2}}{r_{31}r_{32}}\right)
\end{equation}
where $\omega_{1}$, $\omega_{2}$ and $\omega_{3}$ refer to the harmonic vibrational frequencies for Sr-O stretching, bending and O-H stretching modes, respectively. Notice that since the binding energy of the nuclei is $E_{{\rm b}}\sim ka_{{\rm 0}}^{2}=m_{e}e^{4}/\hbar^{2}$ in a molecule and thus the force constant $k$ is proportional to the electron mass $m_{e}$ \cite{demille2015diatomic}, the calculated vibrational frequencies are all proportional to 
\begin{equation}
\omega_{i}\propto\sqrt{\frac{m_{e}}{m_{p}}}=\mu^{-1/2}.
\end{equation}
The stretch-stretch coupling constant $F_{13}$ has been ignored in our calculations since it is less than $1\%$ of the corresponding $F_{11}$ force constant and the use of the diagonal force matrix $\mathbf{F}$ has proven reasonably accurate in the previous work on SrOH \cite{OberlanderThesis}.

\section{Anharmonic vibrations of triatomic molecules}\label{sec:anharmonic-vibrations}

Calculated spectroscopic constants for SrOH indicate that there is a small anharmonic contribution to stretching and bending molecular vibrations as can be seen above. Exact description of the vibrational motion of polyatomic molecules requires inclusion of the anharmonic terms in the molecular potential. A Morse potential of the form \cite{Demtroeder2005} 
\begin{equation}
E_{{\rm Morse}}=E_{{\rm b}}\left[1-{\rm e}^{-a\left(R-R_{{\rm e}}\right)}\right]^{2}
\end{equation}
provides a good approximation for the anharmonic vibrational potential of a diatomic molecule. It can be shown that the vibrational energy levels for a diatomic molecule take the form \cite{Demtroeder2005}
\begin{equation}
E_{v}=\hbar\omega_{0}\left(v+\frac{1}{2}\right)\underbrace{-\frac{\hbar^{2}\omega_{0}^{2}}{4E_{{\rm b}}}}_{x}\left(v+\frac{1}{2}\right)^{2}
\end{equation}
where the $\mu$ constant dependence manifests as $\omega_{0}\propto\mu^{-1/2}$ for the harmonic and $x\propto\mu^{-1}$ for the anharmonic constant. Continuing to treat $m_{e}$ as fixed without loss of generality, we note that the binding energy $E_{{\rm b}}\sim E_{{\rm el}}\sim\frac{e^{2}}{a_{0}},$ where $a_{0}$ is the Bohr radius, does not directly depend on the proton mass $m_{p}$ and is therefore independent of $\mu$ \cite{demille2015diatomic}.

For a polyatomic molecule, local bond stretching vibrations like Sr$\leftrightarrow$O and O$\leftrightarrow$H can also be effectively treated as Morse oscillators \cite{lefebvre2004spectra} and therefore $\omega_{1},\,\omega_{3}\propto\mu^{-1/2}$ and $x_{11},\,x_{33}\propto\mu^{-1}$. For bending vibrations of linear triatomic molecules like SrOH it can also be analytically shown \cite{hirano2018bending} that vibrational levels become $\hbar\left(v+1\right)\sqrt{f/\mu_{{\rm bend}}}$ where $f\propto m_{e}$ is the force constant for the bending motion (see Eq. \ref{eq:Force_matrix}) and $\mu_{{\rm bend}}\propto m_{p}$ is the reduced mass of the bending motion.

The anharmonic constants $x_{11}$, $x_{22}$ and $g_{22}$ for SrOH can be expressed in terms of the force constants and vibrational frequencies as \cite{allen1990systematic} 
\begin{equation}
x_{11}=\frac{1}{16}\phi_{1111}-\frac{1}{16}\sum_{i}\phi_{11i}^{2}\frac{8\omega_{1}^{2}-3\omega_{i}^{2}}{\omega_{i}\left(4\omega_{1}^{2}-\omega_{i}^{2}\right)},\label{eq:x11}
\end{equation}
\begin{equation}
x_{22}=\frac{1}{16}\phi_{2222}-\frac{1}{16}\sum_{i}\phi_{i22}^{2}\frac{8\omega_{2}^{2}-3\omega_{i}^{2}}{\omega_{i}\left(4\omega_{2}^{2}-\omega_{i}^{2}\right)},\label{eq:x22}
\end{equation}
\begin{equation}
g_{22}=-\frac{1}{48}\phi_{2222}-\frac{1}{16}\sum_{i}\phi_{i22}^{2}\frac{\omega_{i}}{4\omega_{2}^{2}-\omega_{i}^{2}}.\label{eq:g22}
\end{equation}
The Morse potential provides a good approximation to bond-stretching motions of linear polyatomic molecules with $x_{11},\,x_{33}\propto\mu^{-1}$. Without loss of generality, consider $m_{e}$ fixed and, therefore, change in $\mu$ corresponds to change in $m_{p}$ \cite{demille2008enhanced}. From the dimensionality comparison of Eq. \ref{eq:x11}, \ref{eq:x22} and \ref{eq:g22} we conclude that $x_{22},\,g_{22}\propto\mu^{-1}$.

\section{Extensions of proposed work}

\subsection{Isotopic substitution}

Vibrational frequencies of the normal modes in polyatomic molecules depend on the constituent atomic isotopes. Strontium has four stable isotopes with atomic masses 88, 86, 87 and 84 and natural abundances of $82.58\%$, $9.86\%$, $7\%$ and $0.56\%$, respectively. Additionally, a deuterated version of the molecule SrOD has been previously experimentally analyzed \cite{anderson1992millimeter}. While for a diatomic molecule the dependence of the molecular vibrational constants on the reduced mass $\mu_{{\rm red}}$ is relatively simple, $\omega\propto\mu_{{\rm red}}^{-1/2}$ and $x\propto\mu_{{\rm red}}^{-1}$, even for a linear triatomic molecule the change in harmonic and anharmonic vibrational constants as a function of isotopic substitution is more complex, as discussed above. While the focus of this paper is on the most abundant \textsuperscript{88}Sr\textsuperscript{16}O\textsuperscript{1}H isotope, potentially other SrOH isotopes could be useful for $\mu$ variation experiments as well.

\subsection{``Frozen'' SrOH}

In order to observe spectral signatures of the resonant absorption of bosonic dark matter previous proposals considered using a pressurized gas container at room temperature with H\textsubscript{2}, O\textsubscript{2}, CO, N\textsubscript{2}, HCl or I\textsubscript{2} \cite{arvanitaki2017resonant} or a cryogenic buffer-gas-cooled sample of O\textsubscript{2} molecules \cite{santamaria2015axion}. Alternatively, one could consider using SrOH molecules embedded in a cryogenic noble-gas matrix. High atomic densities of order 10\textsuperscript{17} cm\textsuperscript{-3} have been demonstrated with spin coherence times approaching $\sim1$ s under some conditions \cite{weinstein2016longitudinal}. Laser spectroscopy of the macroscopic sample of ``frozen'' SrOH could allow probing ALP masses in the $\mu{\rm eV}$ and ${\rm meV}$ range for dark-matter induced rotational and vibrational transitions, respectively. We would like to point out that a similar approach of using diatomic molecules embedded in a solid inert-gas matrix has been proposed for performing EDM experiments with projected $\sim10^{-37}\,e\cdot{\rm cm}$ sensitivity \cite{vutha2018EDM}. However, the approach with frozen polyatomic molecules for dark matter searches does not require the application of MV/cm external electric fields for molecular orientation in the lab frame, thus significantly simplifying experimental design. A more extensive analysis of this approach is beyond the scope of this work.

\end{appendix}

\bibliographystyle{apsrev4-1}
\bibliography{SrOH_UDM_merged}

\end{document}